\documentclass[11pt,final,3p,fleqn,onecolumn]{elsarticle}
 \usepackage{setspace}

\usepackage[bitstream-charter]{mathdesign}
\usepackage[T1]{fontenc}
\usepackage[utf8]{inputenc} 
\usepackage[normalem]{ulem}

\usepackage{physics}

\usepackage{hyperref}

\hypersetup{
    colorlinks=true,
    linkcolor=blue,
    filecolor=magenta,
    urlcolor=cyan,
}
\usepackage{booktabs}
\usepackage{xcolor}
\usepackage{color}

\definecolor{mygreen}{rgb}{0,0.6,0}
\definecolor{mygray}{rgb}{0.5,0.5,0.5}
\definecolor{mymauve}{rgb}{0.58,0,0.82}

\definecolor{bostonuniversityred}{rgb}{0.8, 0.0, 0.0}
\definecolor{blue(pigment)}{rgb}{0.2, 0.2, 0.6}
\definecolor{blue(ncs)}{rgb}{0.0, 0.53, 0.74}
\definecolor{antiquefuchsia}{rgb}{0.57, 0.36, 0.51}

\usepackage{array}
\usepackage{mathtools}
\usepackage{wasysym}
\usepackage{verbatim}
\usepackage{relsize}
\usepackage{multirow}
\usepackage{multicol}
\usepackage{bm}

\usepackage{amsmath}

\def \cand {\mathrm{c}}
\def \neigh {\mathrm{s}}

\usepackage{enumerate}
\usepackage{enumitem}
\setlist{nosep}

\newcommand{\lstin}[1]{{\ttfamily #1}}
\usepackage{listings}
\newcommand*{\tran}{^{\mkern-1.5mu\mathsf{T}}}

\DeclareMathAlphabet{\pazocal}{OMS}{zplm}{m}{n}

\usepackage{lscape}		
\usepackage{fancyvrb}

\usepackage[mathscr]{euscript}

\newcommand{\DD} { {\ensuremath{\Omega}}}  

\newcommand{\xDP}{{\ensuremath{ \textbf{\textit{x}}^\star}}}
\newcommand{\balpha}{{\ensuremath{\pmb{\alpha}}}}

\def \pF  {{\ensuremath{p_{\pazocal{F}}}}} 
\def \pS  {{\ensuremath{p_{\pazocal{S}}}}} 

\usepackage{textcomp}

\usepackage{bigints}

\newcounter{MVremark}

\newcounter{AGremark}

\journal{Structural Safety}

\usepackage{dsfont}

\usepackage[left]{lineno} 

\def \IS {\textsf{IS}}
\def \AS {\textsf{AS}}

\newcommand{\Ns}{{\ensuremath{N_{\mathrm{sim}}}}}
\newcommand{\Nv}{{\ensuremath{N_{\mathrm{var}}}}}
\newcommand{\Nr}{{\ensuremath{N_{\mathrm{run}}}}}

\def \X   {\ensuremath{{\pmb{X}}}} 
\def \x   {\ensuremath{{\pmb{x}}}} 
\def \gX  {\ensuremath{g\left( \X  \right)}} 
\def \gx  {\ensuremath{g( \x )}}

\newcommand{\PF}[1] {\ensuremath{{p_{\pazocal{F}}^{\left(#1\right)}}}}
\newcommand{\PM}[1] {\ensuremath{{p_{\pazocal{M}}^{\left(#1\right)}}}}
\newcommand{\PO}[1] {\ensuremath{{p_{\pazocal{O}}^{\left(#1\right)}}}}

\newcommand{\POann}[1] 
{\ensuremath{{
            p_{\pazocal{O},\mathrm{ann}} 
            ^{\left(#1\right)}
}}}

\newcommand{\pfv[1]}  {{
                \ensuremath{
                    p_{\mathrm{f,v}}^{\left(#1\right)}
                }
            }} %

\newcommand{\pfw[1]}  {{
                \ensuremath{
                    p_{\pazocal{F}, \mathrm{w}}^{\left(#1\right)}
                }
            }} %

\def \pf  {{\ensuremath{p_{\mathrm{f}}}}} 
\def \pF  {{\ensuremath{p_{\pazocal{F}}}}} 
\def \pS  {{\ensuremath{p_{\pazocal{S}}}}} 

\def \IF  {{\ensuremath{\mathbb{1}_{\pazocal{F}}\left(\x \right)}}} 
\def \IFk  {{\ensuremath{\mathbb{1}_{\pazocal{F}}\left(\x_k \right)}}} 

\def \Fexact  {{\ensuremath{\pazocal{F}}}} 
\def \Sexact  {{\ensuremath{\pazocal{S}}}} 
\def \FS      {{\ensuremath{\partial\Fexact}}} 

\newcommand{\Fn}[1]   {{\ensuremath{\pazocal{F}^{(#1)}}}} 
\newcommand{\Sn}[1]   {{\ensuremath{\pazocal{S}^{(#1)}}}} 
\newcommand{\Mn}[1]   {{\ensuremath{\pazocal{M}^{(#1)}}}} 
\newcommand{\On}[1]   {{\ensuremath{\pazocal{O}^{(#1)}}}} 

\def \FN  {{\ensuremath{\pazocal{F}^{(\Ns)}}}} 
\def \SN  {{\ensuremath{\pazocal{S}^{(\Ns)}}}} 
\def \MN  {{\ensuremath{\pazocal{M}^{(\Ns)}}}} 
\def \ON  {{\ensuremath{\pazocal{O}^{(\Ns)}}}} 
\def \IN  {{\ensuremath{\pazocal{I}^{(\Ns)}}}} 

\def \RSp  {{\ensuremath{\mathcal{R}}}} 
\def \RnSp {{\ensuremath{\mathcal{R}^{\mathrm{n}}} }} 
\def \GcSp {{\ensuremath{\mathcal{G}^{\mathrm{c}}} }} 
\def \GSp  {{\ensuremath{\mathcal{G}}}} 
\def \USp  {{\ensuremath{\mathcal{U}}}} 

\usepackage{textcomp}

\begin{document}
\begin{frontmatter}



\title{
    Failure Probability Estimation and Detection of Failure Surfaces   \\
via Adaptive Sequential Decomposition of the Design Domain
}


\author{Aleksei Gerasimov}
\address{Institute of~Structural Mechanics, \\ Brno University of~Technology, Veve\v{r}\'{i} 331/95, 602 00 Brno, Czech Republic}

\author{Miroslav Vo\v{r}echovsk\'{y}}
\address{Corresponding author. \\
Institute of~Structural Mechanics, Brno University of~Technology, \\ Veve\v{r}\'{i} 331/95, 602 00 Brno, Czech Republic,\\
e-mail: vorechovsky.m@vut.cz}


\begin{abstract}

We propose an algorithm for an optimal adaptive selection of points from the design domain of input random variables that are needed for an accurate estimation of failure probability and the determination of the boundary between safe and failure domains. The method is particularly useful when each evaluation of the performance function $g(\x)$ is very expensive and the function can be characterized as either highly nonlinear, noisy, or even discrete-state (e.g., binary). In such cases, only a~limited number of calls is feasible, and gradients of $g(\x)$ cannot be used. The input design domain is progressively segmented by expanding and adaptively refining a~mesh-like lock-free geometrical structure. The proposed triangulation-based approach effectively combines the features of simulation and approximation methods. 
The algorithm performs two independent tasks: (i) the \emph{estimation} of probabilities through an ingenious combination of deterministic cubature rules and the application of the divergence theorem and (ii) the sequential \emph{extension} of the experimental design with new points. The sequential selection of points from the design domain for future evaluation of $g(\x)$ is carried out through a new learning function, which maximizes instantaneous information gain in terms of the probability classification that corresponds to the local region.  The extension may be halted at any time, e.g., when sufficiently accurate estimations are obtained. Due to the use of the exact geometric representation in the input domain, the algorithm is most effective for problems of a~low dimension, not exceeding eight. The method can handle random vectors with correlated non-Gaussian marginals. When the values of the performance function are valid and credible,  the estimation accuracy can be improved by employing a~smooth surrogate model based on the evaluated set of points. 
Finally, we define new factors of global sensitivity to failure based on the entire failure surface weighted by the density of the input random vector.

\end{abstract}

\begin{highlights}
    \item A novel adaptive sequential extension of the sample by building a~mesh-like geometrical structure
 
    \item
    A very small number of evaluations of the expensive model, which may return a~categorical output only
    
    \item
    Sequential design extension optimally balances between exploitation and exploration
    
    \item An accurate estimation of probabilities via the divergence theorem and cubature rules
   
    \item
    Novel global sensitivities to failure based on the entire failure surface, weighted by probability density    
    

    \item
    Robust with respect to multiple designs points and strong non-linearities


\end{highlights}

\end{frontmatter}
\section{Introduction}

Despite the progress made in the second half of the last century \cite{Rackwitz2001}, the current interest of the industry and researchers shows that the issue of determining \emph{failure probability} is still an important one. New methods are still being developed, motivated by the need to solve real, complex problems. Engineering products (be it structures, mechanical systems, infrastructure, and so on) or processes, are nowadays often represented by computationally extensive mathematical models. As noted by \citet{Rackwitz2001}, in practical engineering, there is an interest in \emph{reliability optimization} of products and processes, that is, optimization with reliability constraints \cite{BeySen:CMAME:07}. However, reliability optimization methods encompass reliability analysis and call it repeatedly. 
In the standard setting of the problem, the limit state function (LSF) \gx\ is a~computational model representing the state of the product or process, features $\Nv$ input random variables in vector $\X$ with a~known or assumed joint probability density function $f_{\X}(\x)$. In most cases, the limit state function is formulated in such a~way that negative values of  \gx\ signal failure state, but the failure condition can be formulated differently, too.
There are also problems in which \gx\ is a~discrete state function or the function provides just binary output information: failure or safe operation for a~given combination of inputs \x\ \cite{VORECHOVSKY2022115606}.

The spectrum of existing methods for failure probability estimation is very rich. Their categorization can be done depending on how much they focus on and work with three groups of input information: (i) the geometry and topology of the input space of input variables collected in a~random vector $\X$, (ii) the probability density function of $\X$, be it the true one or some modified sampling density, and (iii) the limit state function \gx\ which characterizes the product's performance. Based on the way that the methods for the estimation of failure probability handle the outputs of the limit state function, two basic groups can accommodate the majority of the existing methods. The first group is represented by \emph{simulation methods} that use only binary information of the output: a~rare event for a~given combination of input parameters either occurs or not, and these estimation methods count the occurrences of the rare event and associate them with weights.  Examples of the basic simulation methods in this group are Monte Carlo sampling, na\"{\i}ve versions Importance Sampling (\IS) \cite{Harbitz,Shinozuka1983}, or asymptotic sampling (\AS) \cite{Bucher2009}. The crude versions of \emph{simulation methods} are reasonably robust with respect to noise, jumps, and all kinds of non-smooth features in the limit state function \gx, but the variance of the estimator is very large
and  explodes in high dimensions in the case of \IS\ and \AS.
Therefore, the number of necessary LSF evaluations is still far too high for computationally demanding computer models. The use of stratified sampling strategies such as Latin Hypercube Sampling,
quasi Monte Carlo sequences 
space filling or low discrepancy designs, etc., may lead to a~minor decrease in the estimator variance, but still, the simulation methods are impractical in cases when a~single evaluation of \gx\ lasts more than a~few seconds.

Another group of methods can be termed \emph{approximation methods} as they use the numerical value of \gx\  to draw conclusions about how much safe/failed the event is and how far it is from the boundary between failure and safe domains in the input space. To give an example, we name the first order reliability method (FORM) and the second-order reliability theory (SORM), i.e., the foundational methods, which are still perhaps the most important results on which modern design codes for engineers are built. 
The development of the mathematical theory of FORM
\cite{Hasofer1974ExactAI,Rackwitz1978} and the second-order reliability theory (SORM)  \cite{Fiessler1979,Hohenbichler1987,BreitungHohen1989,Tvedt1990} with the related FORM/SORM asymptotic approximations for multinormal integrals using Taylor series expansions of the first/second order
\cite{Breitung1984,Breitung1994} was completed in the eighties \cite{DitlMadsen:StructRelMeth:96,MadsenKrenkLind:MethStructSaf:86,MelchBeck:2017}. These methods approximate the shape of the \emph{failure surface}, that is, the boundary between the safe and failure domains. The approximation is performed in regions with a~high probability density $f_{\X}(\x)$, i.e., in design points which are the most central failure points in the Gaussian space of independent input variables. We assume that the failure surface is an $(\Nv-1)$-dimensional object, and it is often illustrated as the zero-valued contour of the limit state function, $\gx$. Indeed, when the failure surface is a~differentiable function and the function is almost linear in the standard normal space, the failure probability $\pF$ has a~simple relationship to the shortest distance $\beta$ from the origin to the failure surface: $\pF \approx \Phi(-\beta)$. 
However, the design points can be hard to find, especially for functions \gx\ with complicated landscapes. The trick is to make use of the gradients of \gx\ and employ a~deterministic gradient-based search \cite{Liu1991,Hasofer1974ExactAI,Rackwitz1978} with the option of changing the starting point or the parameters of the search in order to increase the chance of success. Not only the ``design point search'' needed for FORM and SORM relies on the assumption that the performance function decreases towards the most central failure point. 
Another well-known strategy, which assumes a~connection between the gradient of \gx\ and a~distance to the failure surface, is the famous Subset Simulation (SuS) \cite{Au2001}, which is a~clever stochastic version of the gradient optimization method; see the critical appraisal by \citet{Breitung:19:RESS}. Similarly to the design point search for FORM, such downhill optimization methods may not be successful for global optimization. Repeated runs of SuS may not help if the search is always initiated from the origin, and also the \emph{geometry information is not stored for future use}. The feature of not reusing the data from previous \gx\ calls, and \emph{ignoring the geometry of the problem} in the input domain, is common to many methods, despite the great potential of such data.

The simulation and approximation methods can be combined. For example, \citet{Hohenbichler1988} improved the estimations made by clever geometrical operations in FORM and SORM by \IS. Similarly to FORM and SuS, some sequential adaptive variants of Importance Sampling methods \cite{Bucher1988,Papaio:Papa:Straub:SeqIS:SS:16} are also based on these assumptions regarding the connection between the \gx\ function value and proximity to the failure surface. There are many other clever methods such as line sampling \cite{Schuller2004,deAngelis2015} or directional simulation \cite{Bjerager1988,Nie2000}, which are also using the numerical values of \gx\ to orient themselves in the space of input variables. The idea of extrapolation with a~sequence of modified sampling densities present in Asymptotic Sampling can be seen in an analogy to the methods making extrapolations based on a~sequence of limit state function reformulations \cite{Naess2009,Luo2022}, thus making assumptions about the role of the supposedly smooth landscape of \gx.

A special class of methods, which currently represents perhaps the most active branch of research in the field, combines the robustness of \emph{simulation} methods such as \IS\  with a~low-fidelity approximation of the true computationally demanding limit state function \gx. The idea is to replace the true high-fidelity mathematical model with a~simpler surrogate function (response surface or a~metamodel) which can be evaluated quickly and which can be used to formulate a~binary indicator signaling a~failure event. Indeed, if the evaluation of the surrogate is fast, there is no problem in running millions of evaluations and employ again the rudimentary \emph{statistical} methods based on various arithmetical or weighted averages to estimate the probability integrals. Additionally, the smooth approximation via the surrogate can help with the selection of the location for the next limit state function evaluation (active learning). The problem is making the approximation fast and accurate especially in the high probability regions of failure surfaces. Smart strategies exploit the information from the already analyzed points from the design domain, and they can adaptively refine the surrogate model; see, e.g. \cite{Sundar2016, TeixOconn:AdaptiveMetamodel:Review:SS2021}. The surrogate is approximating the true \gx\ using various smooth functions: smooth polynomial response surfaces, radial basis functions \cite{Li2018,Shi2019}, Kriging  \cite{Echard2011,Wang2022}, Polynomial Chaos Expansions (PCE) \cite{Marelli2018,Zhou2020}, or classifiers such as artificial neural networks \cite{deSantanaGomes2019,Gomes2020,SaraygordAfshari2022}, Support vector regression surrogates (SVM) \cite{Li2006,BOURINET2011343:fourbranch,Bourinet2016,Pan2017,Roy2022}, or nearest neighbor approximations \cite{VORECHOVSKY2022115606}.
The problem is that some of these methods are hard to use in high dimensions (for example PCE), and many depend on the qualified choice of some parameters by the user (be it the correlation function in Kriging or the kernel 
in SVM). 
Many surrogate models are not robust because they may suffer from overfitting, they might not be sufficiently flexible, and they may require too many points in the design domain.
Researchers keep inventing various combinations of methods. An example can be a~combination of Importance Sampling with Kriging \cite{Echard2013}, or SuS with Kriging \cite{Zhang2022-nf}. These combinations can be seen to weigh the assumptions in, of focus on, various pieces of the inputs to the problem. 

The successful application of most of the existing ``approximation'' and surrogate model methods is conditioned by the fact that the real function returns well-behaved (continuous and smooth) output values. 
While this can be fulfilled in many situations, in practice there are also problems in which the output is discontinuous, non-differentiable (contains jumps in function values), noisy, or the output is only binary, or even \gx\ does not return any answer at all for certain combinations of input values (the existence of ``blind spots'') \cite{VORECHOVSKY2022115606}.
This can happen, for example, with nonlinear solvers of complicated phenomena in which jumps or non-smooth behavior is already embedded in the components of the model. In such cases, most of the \emph{approximation} methods described above break down completely. However, \emph{simulation} methods, which are resistant to these complications, cannot be used, because each evaluation of \gx\ is time-consuming.

In this paper, we propose to build a~solid geometrical structure using the small-sample experimental design (ED), i.e., points, in which LSF \gx\ was previously evaluated. This structure serves two tasks: the \emph{estimation} of the failure probability, and a~further \emph{extension} of the ED. We propose new methods to accomplish both these  tasks while reflecting that each evaluation of LSF is very expensive and that the method must extract as much information as possible from the precious data. The methods use only \emph{binary information} about the state of the system \gx, and combine the robustness of the targeted simulation methods with the efficiency of an approximation method.
The geometrical representation divides the input space into (i) the convex hull embracing the ED points and (ii) the remaining ``outside'' territory. The interior of the convex hull is divided into simplices for which the ED points are the  vertices, and this scaffold can be either (i) locally \emph{refined} in the vicinity of the failure surface (exploitation) or (ii) \emph{expanded} by requesting \gx\ evaluation in an \emph{exploratory point} outside the convex hull. 
The \emph{estimation} of probabilities (Sec.~\ref{sec:estimation}) is achieved via analytical (exact)  
and numerical cubature approaches using exact geometrical entities (hyperballs, hyperplanes, and simplices).
The estimated probabilities are used to suggest an optimal \emph{extension} of the ED (Sec.~\ref{sec:extension}), which is, therefore, \emph{adaptive} (based on the current geometrical structure) and \emph{sequential} (one LSF evaluation point is added at a~time). The decision based on probabilities automatically balances between exploitation and exploration to maximize the information gain in terms of classified probability related to the selected point. The sensitivity of failure probability to individual random variables is evaluated using the new factors based on the failure surface proposed in Sec.~\ref{sec:sens}.



\section{An Overview of the Proposed Method} 

\begin{figure}[!tb]
    \centering
    \includegraphics[width=0.4\textwidth]{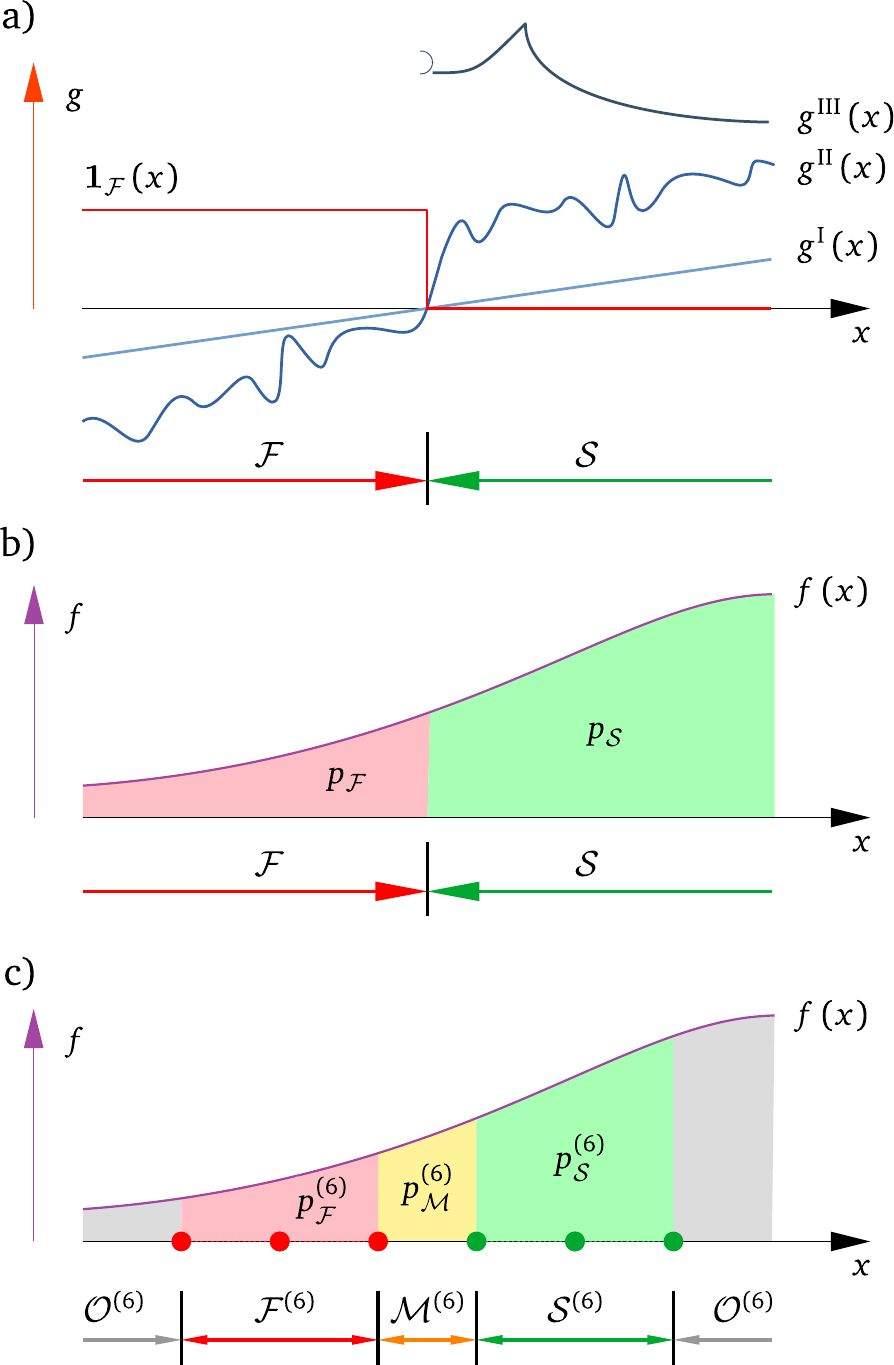}
    \caption{Reliability problem illustrated for 1D. 
        a) three different limit state functions $g(x)$ leading to the same failure domain signaled by the indicator function $\IF$;
        b) probability density $f(x)$ and the geometrical meaning of failure and success probabilities: $\pF$ and \pS, respectively;
        c) interpretation of point-wise information obtained in $\Ns=6$ points by the proposed algorithm: identification of four sets (failure set, safe set, mix set and unexplored set) and the estimation of the corresponding probabilities.
    }
 \label{fig:1DSScheme}
\end{figure}

Consider a~system involving $\Nv$ random variables where  we are interested in the probability of failure, \pF.
%
The random vector is formed by continuous random variables. The $\Nv$-dimensional space at which these random variables are defined is called the \emph{design domain} $\DD$. The probability \pF\  of a~failure event
(exact solution) is defined as the integral of the joint density of random vector \X\ limited to the failure set \Fexact
\begin{align}
    \label{eq:pf:definition1}
    \pF = 
    \idotsint_{\Fexact} f_{\X}(\x) \dd \x
    = 
     \idotsint_{\DD} \IF  f_{\X}(\x)  \dd \x ,
\end{align}
where $f_{\X}(\x)$ and $F_{\X}(\x)$ are the probability density and the cumulative density functions of the random vector \X, respectively.
The failure set $\Fexact \subset \DD$  
is formed by a~union of all regions in the space of all random variables in \X\ in which failure occurs.
The problem is illustrated using a~simple one-dimensional example in Fig.~\ref{fig:1DSScheme}. A~single failure region \Fexact\ extends from minus infinity to the failure surface, which is represented by a~single point here.
The area filled with red color equals \pF\ (compare with Eq.~\eqref{eq:pf:definition1}).
The alternative expression for \pF\ employs an indicator function $\IF$ that returns one for $\x$ falling to the failure set, and zero otherwise.
The indicator function \IF\ may, therefore,  involve the \emph{limit state function} \gx, i.e., a~function that quantifies the performance of the system under investigation. 
This way of expression highlights the dichotomous nature of the problem: the only information from LSF \gx\ needed for the computation of \pF\ is binary (either failure, or success). There are many formulations of LSF leading to the same decomposition of \DD\ into safe and failure domains; see the three versions of the $g$ function depicted in Fig.~\ref{fig:1DSScheme}a. As discussed above, there are, however, many methods which build on the numerical values of \gx, not only the binary information. The values are used to build a~surrogate model, or they serve to detect the steepest descent direction needed for the gradient-based approaches. We argue that relying on the numerical values of \gx\ may be quite problematic: no convergence for highly nonlinear LSFs, reformulation/reparameterization of LSF which retains the failure domain may alter the result of these methods, and discrete-state LSF may break the methods down completely.

When the true decomposition into two disjoint sets is known, $\DD = \Fexact \cup  \Sexact$,  the probability density function $f_{\X}(\x)$ defined in \DD\ can be used to estimate the probabilities; see Fig.~\ref{fig:1DSScheme}b. In practical situations, the domains are not known, but they can be approximated based on the information from the ED. Fig.~\ref{fig:1DSScheme}c illustrates a~situation with $\Ns=6$ ED points for which the classification is known from previous LSF evaluations (green/red color of points signals the safe/failure state). These points can be used to construct an intermediate, temporary approximation of indicator functions by decomposing the design space into four types of domains. 
These are the current
    \emph{failure set} $\Fn{\Ns}$, 
    safe region $\Sn{\Ns}$,
    mixed region $\Mn{\Ns}$, and 
    the remaining ``outside'' territory $\On{\Ns}$,
    which is not belonging to any of the previous three
\begin{equation}
    \label{eq:territories:approx}
   \DD = \underbrace{\FN \cup  \SN \cup  \MN}_{
                   \IN = 
   \text{``inside'' (explored)}} 
                    \quad
                    \cup 
         \underbrace{\ON}_{\text{``outside'' (unexplored)}}.
\end{equation}
We propose constructing this decomposition after each evaluation of LSF.
The existing ED points form a~convex hull (the union denoted as ``inside'' in Eq.~\eqref{eq:territories:approx}), and its interior is segmented into polytopes. We use the simplest polytopes possible in the given space: \emph{simplices}, each having only $\Nv+1$ vertices which are the ED points in which LSF was previously evaluated.
The simplices are built using $\Nv$-dimensional Delaunay ``triangulation''. This ``triangulation'' (generalized into  tetrahedralization in 3D, etc.) decomposes a~convex hull which is tightly embracing all ED points into non-overlapping polyhedra (simplices).
Such a~geometrical approach provides an exact control over the individual simplices; see also Fig.~\ref{fig:integration} left. The interior of each simplex can be temporarily classified as a~part of either 
    $\Fn{\Ns}$, $\Sn{\Ns}$, or $\Mn{\Ns}$. 

We remark that such a~spatial decomposition based on the ED points is performed independently of the way these points were selected (or sampled). The \emph{estimation} of probabilities based on the spatial classification is an optional task, and it can be seen as post-processing of the existing point-wise information. The estimation may or may not affect the future \emph{extension} of ED. The geometrical representation (a~skeleton) can be saved and loaded again if further refinements are deemed necessary.
The described geometrical description allows a~for fast classification of the majority of the input space (leaving only a~negligible portion of probability unexplored) while having accurate representation of the failure surfaces. 
%
The classification proposed in this paper is fast, because we simply classify the simplex interior based on the binary information in the vertices. If all the vertices have the same classification (be it failure of success), the whole simplex inherits this classification. A~simplex with more than one type of classification in the vertices is a~part of \emph{mixed} territory $\Mn{\Ns}$. In this way, the indicator function needed for sampling analysis is defined.

The decomposition uses a~relatively simple computational geometry (Delaunay triangulation \& convex hulls) and together with the described fast classifier, it  independently serves two purposes:
\begin{itemize} \setlength\itemsep{0pt}
   \item 
        the \emph{estimation} of rare event probability, which can be evaluated at any moment of the run via targeted methods proposed in this paper,
    \item 
        a~sequential \emph{extension} of ED, that is, a~selection of the next point \x\ for the evaluation of LSF in such a~way that it delivers the greatest gain in terms of probability.
 \end{itemize}

\begin{figure}[!tb]
    \centering
    \includegraphics[width=0.4\textwidth]{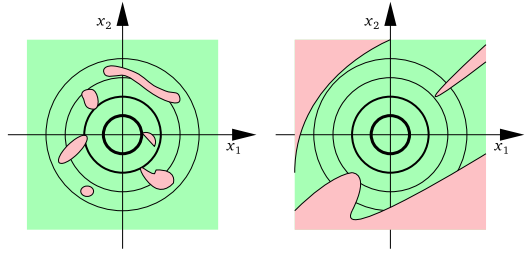}
    \caption{Assumptions about the failure domain.
            Left: closed failure domains for which the proposed algorithm may fail due to a~wrong encapsulation of the failure regions.
            Right: open failure domains fulfilling the assumptions.
    }
 \label{fig:simplex_assumptions}
\end{figure}

Unless stated otherwise, we assume that the failure set is unbounded, meaning that \Fexact\ is an open domain; see~Fig.~\ref{fig:simplex_assumptions} right. For problems with bounded failure regions, i.e., situations where parts of the failure set are scattered or fragmented in \DD\ (see~Fig.~\ref{fig:simplex_assumptions} left), the proposed algorithm is not guaranteed to discover all the closed parts of the failure set. This is because there is a~risk that too small a~failure region will be encapsulated by simplices classified as success events.

\section{The Estimation of Probabilities 
\label{sec:estimation}
}

The skeleton formed by simplices provides, at any stage of the ED extension process, a~rational segmentation of a~compact convex hull into nonoverlapping subdomains. The structured segmentation allows for a~fast and accurate estimation of probabilities. Typically, the convex hull quickly occupies a~large portion of the input space in terms of contained probability. 
The geometrical structure constructed in the inside domain, denoted as $\IN$ Eq.~\eqref{eq:territories:approx}, can be readily used as a~\emph{surrogate model}, which provides an indicator function of the  event types for any point selected from the convex hull. The event types are judged in a~simple way: when all simplex vertices signal the same type of event, then the entire interior of the simplex is associated with it (safe operation or failure). Splitting the probability content of simplices with mixed vertices into probabilities associated with the different events in the vertices is described in Sec.~\ref{sec:mixed}. Finally, the probability contained \emph{outside} the convex hull, region $\ON$, is evaluated, too, giving us important information about how much probability is associated with the yet unexplored territory. In this section, we limit the description of our algorithm to the case of standard jointly Gaussian distribution, i.e., with standard Gaussian density $f_{\X}(\x) \equiv \varphi(\x)$, and the distribution function $F(\x) \equiv \Phi(\x)$, so that the design domain is $\mathbb{R}^{\Nv}$; cases with non-Gaussian vectors and potentially correlated marginals are covered in Sec.~\ref{sec:spaces}.

\begin{figure*}[!tb]
    \centering
    \includegraphics[width=\textwidth]{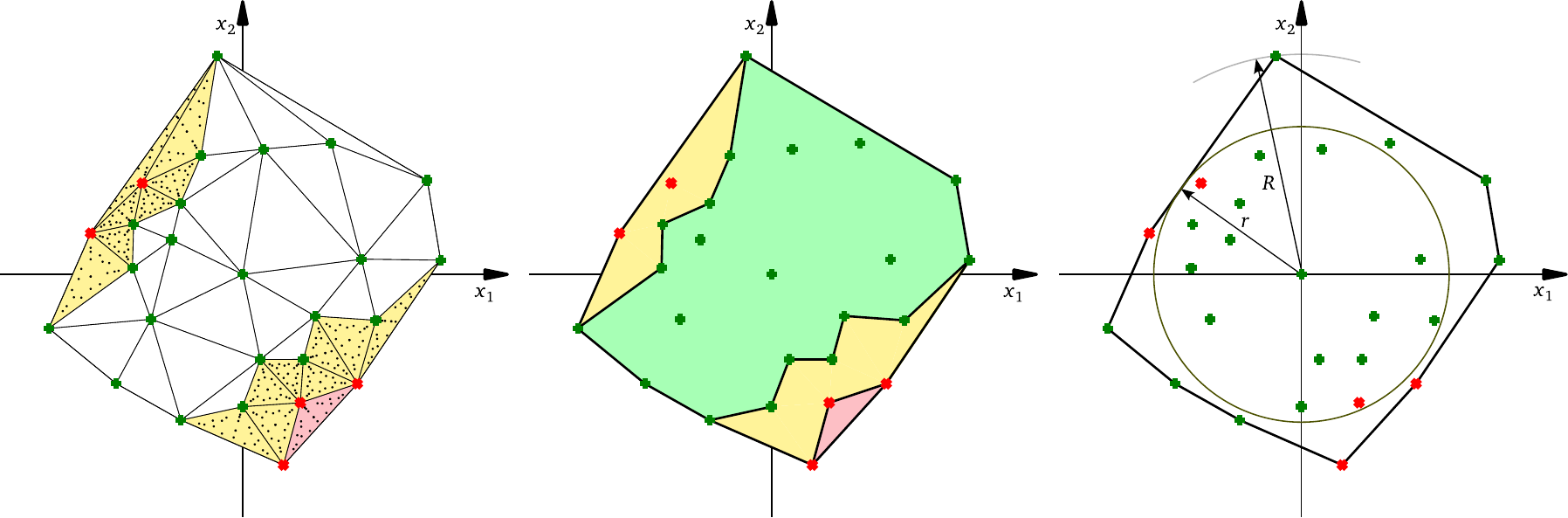}
    \caption{Illustration of probability integration in 2D standard Gaussian space. 
    Left: deterministic cubature of individual simplices with the convex hull.
    Middle: clusters of simplices of the same type with piecewise linear boundaries.
    Right: boundary of the convex hull found between two hyperballs having radii $r$ and $R$.}
 \label{fig:integration}
\end{figure*}

The inside territory consists of three types of domains: $\IN  =  \FN \cup  \SN \cup  \MN $, and each of them is represented by a~set of simplices. Fig.~\ref{fig:integration} shows a~2D example of an ED with 21 safe outcomes (green points) and five failure events (red crosses). It also illustrates the corresponding convex hull and its division into the three territory types:
    the figure on the left highlights the mixed and failure simplices, 
    the middle figure shows clusters of simplices of the same type, separated by the (hyper)planar faces (ridges).
    The figure on the right illustrates that the outside territory is associated with a~probability which is greater than the probability outside the circumscribed \Nv-ball with radius $R$ (the most distant ED point) and, at the same time,     
    smaller than the probability outside the inscribed \Nv-ball with radius $r$ (the nearest facet of the convex hull). These bounds are simple to compute analytically, because the radii $r$ and $R$ are available from the convex hull (see \ref{appendix:nvball}). 
    The solution is based on the fact that a~random distance in the standard Gaussian space has an $\chi$ distribution with \Nv\ degrees of freedom. Eq.~\eqref{eq:prob:out} provides the simple analytical solution for the probability outside the circumscribed \Nv-ball
\begin{equation}
    p_{R,\mathrm{out}} (R;\Nv)  = 1 - F_{\chi} (R; \Nv). 
\end{equation}
To obtain an upper bound on the unexplored probability, one can add the probability in the annulus 
between the \Nv-balls with radii $R$ and $r$: $F_{\chi} (R; \Nv) - F_{\chi} (r; \Nv)$.
The smaller radius~$r$ of the inscribed \Nv-ball is the distance from 
the most central facet point
to the origin, which is known from the convex hull geometry; see \ref{appendix:outside:convex_hull}. 

Apart from the bounds on the outside probability  $\PO{\Ns}$, we propose using integration employing the divergence (Gauss-Ostrogradsky) theorem. The exact geometrical description of the convex hull via a~finite set of (hyper)planar facets makes it very easy to integrate a dot product of the normals with a~wisely selected vector field only along its boundary, instead of computing a~cubature inside the whole convex hull. This procedure decreases the dimension integration domain by one (over the embracing surface instead of the interior volume), and it is described in \ref{sec:divergenceTheorem}.

The same procedure can be used to compute the cubatures of clusters of simplices visualized in Fig.~\ref{fig:integration} middle. It suffices to know the boundaries of clusters of simplices which are of the same event type and which are formed by pairs of neighboring simplices sharing their $(\Nv-1)$-dimensional faces. The analysis of \emph{adjacency} and clustering is very easy to achieve by using the \lstin{QHull}  library, which provides a~list of neighbor simplices for every simplex.

\subsection{The Integration of Individual Simplices \textbf{Inside} the Convex Hull
    \label{sec:estimation:inside}
}

Since we are interested in the rare event probability, that is, the failure probability, we are interested in estimating the probabilities corresponding to the current failure set, $\FN$, and also the mixed set, $\MN$, a~part of which can contribute to failure probability, albeit with an unknown probability share.  
Simplices belonging to the failure set can typically be formed only after the formation of the mixed simplices, and they may not form a~compact set of neighboring failure simplices. The initially scattered failure (red) simplices must be integrated one by one. Moreover, the probability content of individual mixed simplices with uncertain contributions to the rare event can be used to select the best location for an ED extension via refinement, see Sec.~\ref{sec:extension}. Therefore, we present a~tool for the cubatures of individual simplices.

Suppose the convex hull is divided into simplices constructed with $\Ns$ ED points (vertices).
Let us index all existing simplices by $j = 1, \ldots, N_J$ and denote the probability content of the $j$th simplex by $p_j$. 
We propose to 
focus only on the failure and mixed simplices and compute the probability estimation as the sums of probabilities over all simplices sharing the corresponding type
\begin{align}
     \label{eq:PFM}
     \PF{\Ns} &= \sum_{j \in \FN} p_j, \\
     \PM{\Ns} &= \sum_{j \in \MN} p_j.
\end{align}
The probability content of safe (green) simplices should \emph{not} be performed by the analogous sum over individual simplices. The reason is that the probability is typically large, and numerical errors can accumulate and degrade the estimation. The safe probability is better estimated either by (i) subtracting the ``mixed'', ``failure'' and also the ``outside'' probabilities from the total unit probability (see Sec.~\ref{sec:estimation:outside}), or by (ii) grouping the safe simplices, which typically form a solid body, and computing the content inside the group using the cluster boundary via the application of the divergence theorem, see \ref{sec:divergenceTheorem}.

There are many ways of estimating the cubatures in individual simplices, including the Monte Carlo type of integration. In this paper, we introduce a~deterministic cubature based on predefined integration nodes. Consider the $j$th simplex which occupies a~region denoted $\Omega_j$. The target is to estimate the integral
 \begin{equation}
    p_j = \int_{\Omega_j} f_{\X}(\x)  \; \mathrm{d}\x.
\end{equation}
An analytical solution is not known even for the particular case of Gaussian distribution, and only relatively tight bounds for probability of two intersecting hyperplanes were derived in the past \cite{Ditlevsen1979}. However, these bounds are not practical in the present case, where there can be a~large number of simplices, each representing many such intersections. We propose usign deterministic cubature rules which provide exact solutions for polynomials up to a~given degree, and whose application is fast and effective.
The cubature integration involves a~multiplication of simplex volume $V_j$ by a~weighted average value of the integrand $f_X(\x)$. 
The volume is calculated using the determinant of coordinate matrix
\begin{equation}
     V_j = \frac{1}{\Nv!} 
      \begin{vmatrix}
     \x_2 - \x_1\\ 
     \x_3 - \x_1\\
     \vdots \\
     \x_{\Nv+1} - \x_1
 \end{vmatrix}
 ,
\end{equation}
where $\x_k$ are the vertices of simplex $j$, $k=1,\ldots,\Nv+1$.
The weighted average is represented by densities evaluated in $n_l$ predefined integration nodes $\x_l$, each multiplied by the known weights $w_l$. Finally, the probability content is estimated via
\begin{equation}
    p_j = V_j \sum_{l=1}^{n_l} f_{\X}\left(\x_l\right)w_l.
\end{equation}
The coordinates $\x_l$ of the integration nodes inside the simplex are predefined in a~unit simplex and can be easily transformed to the rotated and stretched simplex $\Omega_j$. The weights sum to unity: $\sum_{l=1}^{n_l}w_l = 1$. Some weights may not be positive as many cubature schemes provide integration weights with both signs. Many integration schemes were developed in the last century. The first integration schemes for simplices in an arbitrary dimension were developed by Hammer and Stroud~\cite{10.2307/2002484, 10.2307/2002945}. Later, \citet{silvester} proposed two basic variants of cubatures: open (integration nodes are strictly inside the simplex volume) and closed (integration nodes can be placed on simplex facets). \citet{Grundmann1978} developed general degree schemes for $n$-simplex and proved that the
transformation into unit simplex is unique and invariant. Their scheme is widely used nowadays. 
\citet{DELOERA2013232} also contributed to new efficient cubature via recursive formulas, and the development of new ones continues until this day, see \cite{LIN2020125140, isaac2020recursive, tommaso_zerroudi_2020}.


In our numerical studies, we employed a~free software package for the \lstin{Python} environment named \lstin{quadpy}, available at  \href{https://zenodo.org/record/4340036}{https://pypi.org/project/quadpy/} 
\cite{quadpy:SW}.
The library implements, among others, a~hierarchical refinement method for the integration of $n$-dimensional simplices by \citet{Grundmann1978}. Fig.~\ref{fig:integration} left shows the example of one particular set of integration nodes in mixed and failure simplices.

We remark that these deterministic cubatures have drawbacks when applied to integrating probability densities. These densities are typically not the polynomials for which the node positions \& weights have been derived and, therefore, can be neither exactly approximated, nor ``covered'' by a~polynomial of a~sufficiently high degree due to Runge's phenomenon.  
During the aggressive exploratory phase in high dimensions, simplices can span regions with a~high local density as well as remote regions with an extremely low density. There is a~risk of obtaining negative cubature despite the fact that all the densities are positive. This occurs due to the fact that some integration rules use negative weights associated with some integration nodes, and this can lead to dubious integration results. When such behavior is detected, another integration technique, using adaptive refinement with automatic error evaluation, is desirable. Since individual simplices can be integrated independently, the task is ideal for parallel processing.

By adding a~new point to the ED, the triangulation is typically changed only locally and, therefore, it is not hard to keep track of the triangulation and the estimations: 
invalidated simplices are removed from a~database and only newly added simplices are integrated.


\subsubsection{Mixed Simplices}
\label{sec:mixed}

The sum of probabilities computed for ``mixed'' simplices, \PM{\Ns}, may represent a~non-negligible value, and especially for small EDs may easily be greater than the failure probability $\PF{\Ns}$ estimated using the purely ``failure'' simplices. It is reasonable to expect that the boundary between failure and safe domains passes through mixed simplices. A~question arises as to how the probability content $p_{j}$ of simplex $j$ obtained previously via selected cubature should be split into failure and safe parts so that the ``failure share'' can be added to the failure probability estimate.

\begin{figure}[!tb]
    \centering
    \includegraphics[width=5cm]{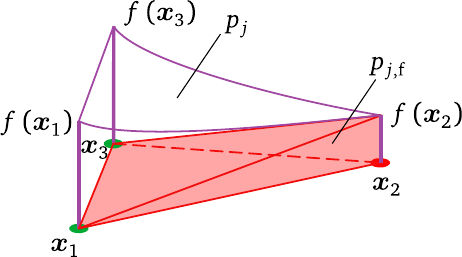}
    \caption{Two-dimensional illustration of the problem of ``mixed'' probability assignment of a~single simplex into ``safe'' and ``failure'' probabilities.
    }
 \label{fig:cakes}
\end{figure}

We propose a~simple 
estimation of failure share in simplices by a~linear approximation of failure density over the simplex. 
The following formula involves probability densities at failure vertices only and yields the exact solution~\cite{Lauffer1955} for the probability content of a~planar failure body; see the two-dimensional illustration in Fig.~\ref{fig:cakes}   
\begin{equation}
 \label{eq:failureinmixed}
    p_{j, \mathrm{f}} = V_j  \frac{  \sum_{k=1}^{\Nv+1}
                                      \IFk  
                                      \; 
                                      f_{\X}(\x_k) 
                                  } 
                                  {\Nv+1},
\end{equation}
where \IFk\ is the indicator function applied to all $\Nv+1$ vertices $\x_k$ indexed by $k$, in which LSF outcome is known. 
We remark that in an even simpler alternative, one can simply multiply the simple probability $p_j$ by the number of failed vertices over the number of all $\Nv+1$ vertices, but this crude alternative may be less accurate than Eq.~\eqref{eq:failureinmixed}.

Once the probability content of each mixed simplex is distributed into failure and success probabilities, the current estimation of failure probability can be obtained by simply summing the full contributions of all purely failure simplices with the sum of the ``failure'' contributions from all the mixed simplices.

Finally, the current estimation of failure probability is the sum of the probability contents in the purely failure simplices with the sum of failure shares in the mixed simplices
\begin{align}
\label{eq:vertex2}
 p_{\mathrm{f,v}} =
         \sum_{j \in \FN} p_j
        +
        \sum_{j \in \MN}  p_{j, \mathrm{f}} \, .
\end{align}
The running estimations of the probabilities can be used to formulate a~\emph{stopping criterion}. One can be tempted to halt the ED extension and LSF evaluation once 
    (i)   the estimation of failure probability becomes stable (varying within narrow margins), or when 
    (ii)  the sum of the estimated mixed and outside probabilities becomes less than the failure probability.
However, none of these criteria alone guarantee either that there is no failure region yet to be discovered, or that the existing approximation of the failure surface is refined sufficiently. However, as will be demonstrated using numerical examples, when the basic requirement of open failure domains is met, the method is robust in the sense that the estimations tend to the correct solution as new points are being added to ED.

\subsection{The Integration of Probability \textbf{Outside} the Convex Hull
    \label{sec:estimation:outside}
}

Once the number of points in ED exceeds $\Nv+1$, multiple simplices can be constructed and their number increases rapidly upon adding new points. As illustrated in Figs.~\ref{fig:integration} and \ref{fig:extension}, the set of simplices forms a~convex hull whose probability content rapidly approaches one as the points are being added to ED. In order to calculate what probability is inside the convex hull accurately, it is better to calculate the content in the entire set of simplices rather than summing the estimates of the cubatures of individual simplices. 
The outer linear boundaries (line segments, planes, hyperplanes) are precisely described and the use of the \emph{divergence theorem}, which limits the integration to the $\Nv-1$ dimensional boundary, is an excellent tool for the calculation. \ref{sec:divergenceTheorem} is devoted to the proposed application of the divergence theorem for the probability integration inside closed regions as well as the integration of the uncontained part of the Gaussian density  outside the convex hull. These estimations can be elegantly achieved by suitable choices of the vector fields in the divergence theorem, see \ref{sec:divergenceTheorem}.

\section{Extension of the Experimental Design
    \label{sec:extension}
}
One of the default assumptions behind the development of the proposed algorithm for expensive functions, even if it is just binary LSFs, is that the spatial information from the existing ED points, including their arrangement, must be utilized effectively. Throwing ED points in the style of Monte Carlo methods (Importance sampling, Subset simulation, Asymptotic sampling, Line sampling, etc.) is not effective, and it is even impossible in cases when each single evaluation of LSF is very costly in terms of time or resources. The ability of \emph{sequential extension} calls for the definition of a~\emph{criterion} (learning function) employed in the selection process among the proposed ED point candidates. The criterion should be designed to balance between \emph{exploration} (discoveries and annotation of new territories) and \emph{exploitation} (refinement of the local information about promising/important neighborhoods, i.e., mixed simplices).

No two ED points should be requested close to each other unless they deliver a~significant gain in terms of probability classification (\pF\ estimation). 
The gain in terms of probability can be obtained either by
    (i) the \emph{refinement} of approximation of the failure surface via exploitation or 
    (ii) the \emph{expansion} of the territory with a~new region for which an approximation of \gx\ does not exist yet (exploration). 
The constructed geometrical structure based on the ED points can be effectively used to support the decision about the new ED point location. The space is decomposed into the
   (i) closed region $\IN$, which is fragmented into individual simplices inside the convex hull, and the 
   (ii) outside region $\ON$ spanning to infinity, which can also be fragmented into $n_o$ subregions by using the convex hull.
The decomposition of the outside region is dictated by the  $n_o$ walls enclosing the convex hull via the boundary facets (walls). The rays connecting the origin of the coordinate system with the outer ED points form the edges being parts of boundaries between $n_o$ outer subregions, which span to infinity.

\begin{figure*}[!tb]
    \centering
    \includegraphics[width=\textwidth]{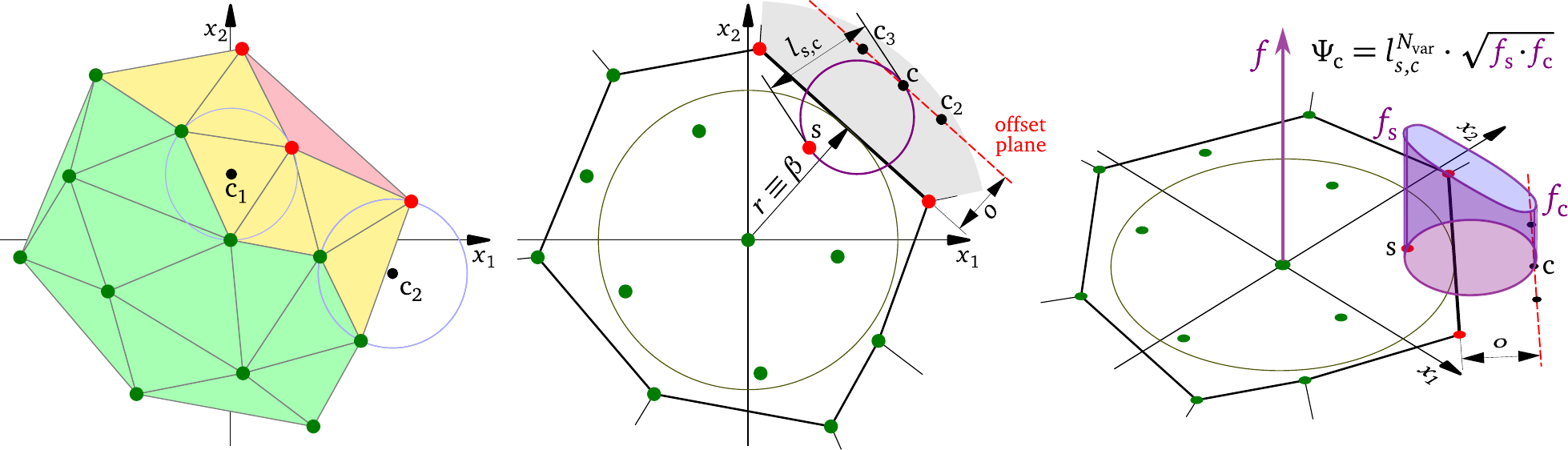}
    \caption{
            Left: break mixed simplex;
            Middle: break outside simplex spanning to infinity.
            Right: selection of the exploration candidate via the $\Psi_c$ criterion.
    }
    \label{fig:extension}
\end{figure*}

The unicolor simplices (be it the failure or safe regions) are not considered in the selection process. To improve the territorial classification, we select the new ED point to break either 
  (1) the ``critical'' \emph{mixed} simplex or 
  (2) one of the $n_o$ outside regions; see Fig.~\ref{fig:extension}.  
We propose a simple rule for the selection between the two, which is based on probability estimation. From all the mixed simplices, we select the one with the greatest probability content $p_j$. Breaking this simplex into more, smaller regions and locally regenerating the spatial decomposition will improve the classification in the vicinity of the true failure surface, preferably in the high probability density region.
At the same time, if the unexplored probability outside the convex hull is large, it also makes sense to insert a~new exploratory ED point outside the convex hull and to select the heaviest of the $n_o$ regions. We formulate a simple rule: (1) when the probability $\max_{j \in \pazocal{M}} p_j$ in the heaviest \emph{mixed} simplex $j$ is greater than the outside probability found behind one critical facet, the next ED point is of the exploitation type and its selection is based on the $j$th mixed simplex. Otherwise, (2) we select the new ED point from the outside domain with the argument that it makes more sense to decrease the unexplored probability. In order to make the comparison of probabilities, we approximate the outside probability behind the critical facet as $p_{\beta} = \Phi ( - r)$, which is a~FORM-style probability based on the probability in Gaussian space found behind the hyperplane  at the distance $r$ from the origin ($ \Phi$ is 
the cumulative distribution function of standard Gaussian variable). The current convex hull is formed by a~set of bounding walls (hyperplanes) and we first select the hyperplane with the shortest distance $r$ from the origin and split the unit probability content by this plane; see the illustration in Fig.~\ref{fig:extension} middle, in which the critical wall is in the first quadrant.

In scenario (1), $\max_{j \in \pazocal{M}} p_j > p_{\beta}$, the motivation is to break the heaviest mixed simplex found inside the convex hull. The new ED point is selected as the center of the \Nv-ball circumscribed about the $\Nv+1$ vertices of the $j$th simplex. When the simplex has a~regular shape, the point appears inside the critical simplex; see candidate $c_1$ in Fig.~\ref{fig:extension} left. However, it may happen that elongated polyhedron produces a~point lying outside its volume; see candidate $c_2$  in  Fig.~\ref{fig:extension} left.
In any case, this center point will break the heaviest mixed simplex $j$ and improve the regularity of the refinement in the region close to the true failure surface (see also the minimax space-filling criterion \cite{JohMooYlv:MixiMinMiniMax:JSPI:90,Pronzato:MinimAndMaxim:17,EliVorSad:miniMax:ADES:20}).

In scenario (2),  $\max_{j \in \pazocal{M}} p_j < p_{\beta}$, we select the most neglected direction of the convex envelope expansion, which is dictated by the hyperplane with the shortest distance $r\equiv \beta$ to the origin. The next ED point is selected from the region behind this wall; see Fig.~\ref{fig:extension} middle. The exact position of the point is selected from many randomly generated candidates placed on a~plane parallel to the nearest wall with a~given offset $o$. The candidates can be generated by projecting a~random sample onto the plane. From all the candidates, we select the one which maximizes the $\Psi_c$ criterion introduced in Eq.~(27) of \cite{VORECHOVSKY2022115606}. 
The value of the criterion for any candidate is computed as 
$\Psi_c = \sqrt{f_\cand \, f_\neigh} \cdot  l_{\neigh,\cand}^{\Nv} $, where the first term is the geometrical average of the probability density in candidate $c$ and its nearest neighbor $s$, and the second term features the distance between them. Raising this distance to the domain dimension quantifies the volume between them  \cite{VORECHOVSKY2022115606}. The maximization of this criterion selects a~candidate which maximizes the rough probability content of a region between the candidate and its nearest existing ED point; see Fig.~\ref{fig:extension} right. In this way, the selected candidate is found far from the existing point and at the same time prefers a higher probability density region. Yet, its selection from a~pool of randomly generated candidates guarantees a~sort of organic shape and prevents the formation of unwanted patterns.

The only remaining question is the choice of the offset $o$. Its value should depend on the user's decision regarding the required aggressiveness of exploration. Too small an offset will lead to a~consumption of many points for a~given ball radius, but the search directions will be densely populated. Unfortunately, the discovery of the first failure point in real applications with low failure probabilities (distant failure surfaces) may cost many LSF evaluations. On the other hand, an overly aggressive exploratory phase may place ED points at too long distances from the central mean value. Clearly, the selection of an optimal offset depends not only on the outside probability itself, but also on the problem dimension. In high dimensions of the input domain, a~small addition of radial distance to the important ring at radius $\Nv-1$ leads to sudden occupancy of a~very large probability content, see \ref{appendix:nvball} and also  \cite{VORECHOVSKY2022115606}. We have found that a reasonable choice valid also in the very early stages of the exploratory phase (very few LSF evaluations and no failure points yet) can be recommended as
        
\begin{equation}
 \label{eq:explor:offset}
    o =S_{\chi}^{-1}\left(\frac{\PO{\Ns}}{\mathrm{Sur}\left[  B_{R+q} \right]}; \Nv\right)    - r,
\end{equation}
where $S_{\chi}^{-1}(\cdot; \Nv) $ is the inverse survival function of the $\chi$ distribution, used to compute the radius of $B_{\rho}$ with the exterior of the prescribed probability, see Eq.~\eqref{eq:invSchi}, and  $\mathrm{Sur}\left[  B_{R+q} \right]$  is the surface of \Nv-ball the with radius of $R+q$, where $R$ is the distance from the most distant ED point from the origin (the radius of a~circumscribed ball from Fig.~\ref{fig:integration} right) and  $q=1$. The offset $o$ is a~free parameter of the proposed method, and by selecting a~particular rule for $o$, the user can control the aggressiveness of the exploration. The presented Eq.~\eqref{eq:explor:offset} is one of many options which, in our case, showed a~good behavior in the problems of various dimensions and for various types of LSFs. All the numerical examples presented below are computed with this rule.

We remark that in typical applications, failure is a~rare event, and all the ED points added in the extension process initially signal a~safe state. During this \emph{exploratory phase}, no triangulation of the convex hull is necessary. However, once the first rare event occurs in the ED, the triangulation described above is constructed and all four types of domains listed in Eq.~\eqref{eq:territories:approx} can appear.

The described method for ED extension is \emph{adaptive} and \emph{sequential}, i.e., the geometrical representation adapts to the current knowledge (information from the new ED point may locally lead to a~substantial reconstruction of the geometrical skeleton) and the ED points are added one by one. 



\section{Stability, Repeatability, Convergence and Efficiency of the Estimation
    \label{sec:convergence}
}

There are multiple sources of variability in the presented estimations of probabilities. First of all, the extension algorithm builds a~skeleton which partitions the input space, and this decomposition is not repeatable due to the randomness involved in the point selection during the exploratory steps. Therefore, repeated runs of the extension algorithm result in different histories of the space decomposition. Moreover, there are more than one deterministic cubature rules applied to the simplices, and the use of adaptive refinement of the quadratures generally depends on the parameters specified by the user.
Despite the existence of the described sources of differences in estimated probabilities, our experience with countless numerical simulations is that the estimator variability is very low.

The estimators do not provide monotonic convergence towards exact results. The reason is that the geometrical structure is very rough when the ED is small, and the estimated values may then show quite serrated profiles for an increasing number of LSF evaluations, $\Ns$.

The average convergence rates are strongly dependent on the problem dimension, $\Nv$. The reason is twofold. In the initial \emph{exploratory phase}, there is only one type of LSF outcome and the rate with which the convex hull occupies the space is lower for higher dimensions. Therefore, also the decrease in the outside probability estimation is slower in high dimensions. 
Once the first rare event is discovered, the extension algorithm automatically switches between the local \emph{refinement} and global \emph{expansion} of the simplex structure.  However, the proportion of local exploitation steps heavily depends on the shape of the failure surface. If the boundary has only a~small extent in a~high probability region, its refinement is fast. However, if the boundary is close to a~surface of an \Nv-ball in the standard Gaussian space, its extent is large and, therefore, its refinement consumes many ED points. Subsequently, the convergence rate of the estimator becomes low.

The proposed exact geometric approach is limited to small dimensions (not exceeding eight). Beyond this limit, the computational geometry becomes too CPU and/or memory expensive. This is the price paid for having the possibility to work with highly nonlinear, noisy or even binary LSFs, i.e., scenarios in which most of the existing methods either break down (approximation methods and smooth surrogates) or require far too many LSF evaluations (simulation methods).

Compared to other existing methods, the proposed method provides quite reliable and robust estimations obtained for quite low numbers of LSF evaluations. This is true despite the fact that our method uses only binary information from the LSF evaluation: either failure, or success. 

Regarding the computational efficiency of the method, we remark that the \emph{estimation} of probabilities as well as  the selection of the best candidate location (\emph{extension} of ED) can be pre-computed in advance for all potential outcomes while the expensive limit state function is still being computed. In this way, the wall time spent with the proposed algorithm can be \emph{de facto} diminished or even decreased to zero.

\section{Failure Probability Sensitivity to Random Variables 
\label{sec:sens}
}

In a~practical use of methods for the estimation of failure probability, a~question appears as to how many individual variables contribute to the failure event(s). More generally speaking, there is a need to optimize a~product or a~process by analyzing the importance shares of individual variables of a~defined event. Doing so provides us with important information needed to focus on the critical uncertainties and to disregard the unimportant ones to reduce the problem dimension. 

\begin{figure*}[!tb]
    \centering
    \includegraphics[width=\linewidth]{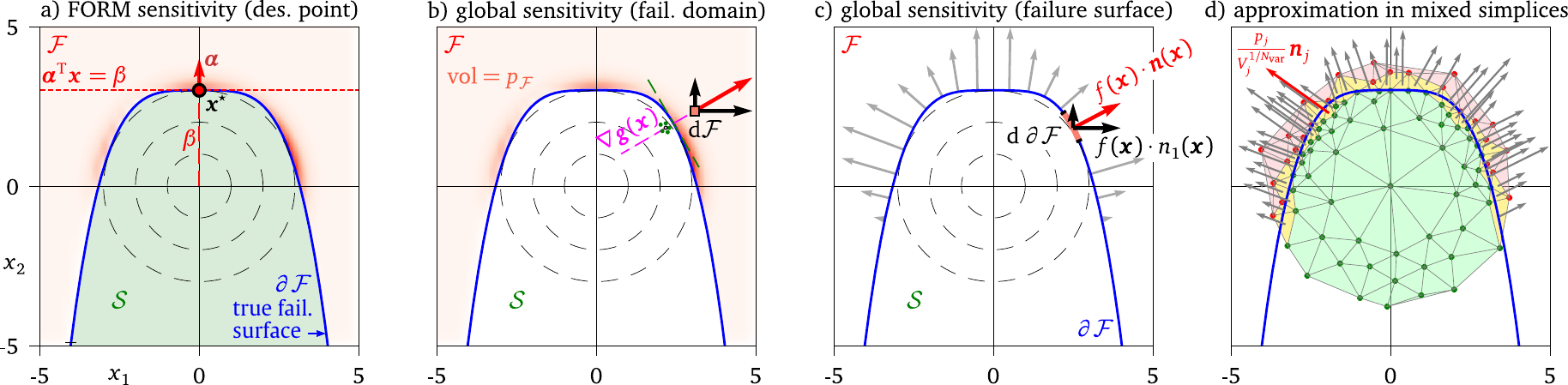}
    \caption{
            Sensitivity measures. 
            a) FORM sensitivities $\balpha = \{ 0, 1\} $ based on a single design point;
            b) sensitivity from  \cite{VORECHOVSKY2022115606}, which is based on the integral over the whole failure domain and the local directions of gradient (approximated via the nearest boundary for binary/discrete LSF);
            c) the proposed definition of sensitivities based on the contour of the failure surface $\FS$; 
            d) illustration of the proposed numerical estimation based on 48 mixed simplices constructed after the evaluation of LSF in 85 ED points.
    }
    \label{fig:sensitivity}
\end{figure*}

When the LSF \gx\ can be reasonably approximated by a~linear function, the FORM solution is very effective not only for the estimation of failure probability, but the by-product of the design point search is the simple evaluation of $\alpha$-sensitivities \cite{HohenRackw:Sens:86,Madsen1988}.
These values have a~clear geometrical meaning, and they have an immediate application in the partial safety factor method in design codes.
Suppose we have a~linear performance function \gx{} with one distinct design point
    $\xDP = \arg\,\min \{ \lVert \x \rVert \; | \; \gx = 0 \}$.
The Euclidean distance of this most central failure point to the origin of the standard Gaussian space is the safety index
    $\beta = \lVert \xDP \rVert
           = \sqrt{ \sum_{v=1}^{\Nv} (x_v^{\star})^2 }
    $.
The distance can also be written using the vector of $\alpha$-sensitivities, $\balpha$, which has the unit size:
$
 \lVert \balpha \rVert
    = \sqrt{ \sum_{v=1}^{\Nv} \alpha_v^2 }
    = 1
$.
Vector $\balpha$ is the negative unit gradient of the performance function at the origin of the standard normal space, i.e., $\balpha$ points in the \emph{important direction} (see the illustration in Fig.~\ref{fig:sensitivity}a; and the coordinates of the design point can be obtained as $\xDP = \beta \balpha$. The components of $\balpha$ (called $\alpha$-sensitivities factors) are regarded as measures of the sensitivity of $\beta$ to the changes in the value of \x\ at the design point (the most central failure point).
Fig.~\ref{fig:sensitivity}a illustrates that in the case of nonlinear LSF the sensitivity based on a single design point may not give the full picture about all the variables  which contribute to failure. The figure depicts the failure surface corresponding to the following LSF. Failure occurs when $\gx \leq 0$ in
 $\gx = \beta - x_1^4/c - x_2$, in which $\beta = 3$ and $c=33$. The use of $c>32$ leads to a~unique design point,
 $\xDP=\{0,3\}$, but there are two additional symmetrically distributed regions around points
  $x_1 = \pm 2.84$ and $x_2=1.02$, which also considerably contribute to the failure probability; see the dark red regions in Fig.~\ref{fig:sensitivity}a.
The FORM $\alpha$-sensitivities based on $\xDP$ are $\balpha = \{ 0, 1\}$ suggesting no sensitivity on the variable $x_1$. Indeed, $\gx$ is linear in $x_2$ in the vicinity of $\xDP$ and does not depend on the zero $x_2$.  
 The failure surface \FS\ is linear at $\xDP$, and the SORM correction fails, because the curvature in the $\beta$-point vanishes. However, the failure event is almost equally sensitive to both variables $x_1$ and $x_2$ \cite{VORECHOVSKY2022115606}.

The recent work of  \citet{VORECHOVSKY2022115606} generalized the split of coordinate contributions to account for the whole failure domain \Fexact, not just the design point. This generalization allows for the consideration of non-linear LSFs and their gradients inside the failure set; see Fig.~\ref{fig:sensitivity}b. The idea is that while any region inside \Fexact\ contributes to the failure probability proportionally to the local density, this local contribution can be split into $\Nv$ contributions using the generalized $\alpha$-factors. In particular, the local contribution of the variable $v$ at a~point $\x$ to a~failure event  \Fexact\ reads
\begin{align}
 \label{eq:loc:contrib}
    s_v^2(\x)
    =  \IF  f_{\X}(\x) \alpha^2_v(\x),
\end{align}
where $\alpha_v(\x)$ is the local contribution of the variable $v$ standardized in such a~way that $\sum_{v=1}^\Nv \alpha_v^2 (\x) = 1$ for any point $\x$.
Summation over all variables yields the complete local contribution to failure probability (density)
\begin{align}
 \label{eq:loc:contrib:allvars}
    \sum_{v=1}^\Nv s_v^2(\x)
    =  \IF  f_{\X}(\x)  \sum_{v=1}^\Nv \alpha^2_v(\x) = \IF  f_{\X}(\x).
\end{align}
The integral of this density over the whole design domain $\pF = \int_{\DD} \IF  f_{\X}(\x)  \dd \x$ is the global failure probability; compare with Eq.~\eqref{eq:pf:definition1}. However, the split into individual variables provided in Eq.~\eqref{eq:loc:contrib} allows for a straightforward definition of the overall contribution of the variable $v$, denoted as $s_v^2(\x)$
\begin{align} 
    s_v^2 
    =  \int_{\DD} s_v^2(\x)  \dd \x
    = \int_{\DD} \IF \,  f_{\X}(\x) \, \alpha^2_v(\x) \, \dd \x
    = \int_{\Fexact}     f_{\X}(\x) \, \alpha^2_v(\x) \, \dd \x.
\end{align}
The question remains what are the local $\alpha$-factors needed to determine the contributions of individual variables. One option would be to use the standardized local gradient of the limit state function
 \begin{align}   
 \label{eq:alpha:grad}
    \alpha^2_v(\x)
    =
        \left(
            \frac{ \nabla g_v(\x)   }
                 { \lVert \nabla \pmb{g}(\x) \lVert}
        \right)^2
    ,
    \quad
    \text{i.e.}
    \;
    \sum_{v=1}^{\Nv} \alpha_v^2(\x) = 1
    \;\; \text{for any} \; \x
    .
\end{align}
Consequently, the sum of all sensitivity measures is equal to one, $\sum_{v=1}^\Nv s_v^2$, and they can be seen as global contributions to the complete failure probability. However, taking the gradient of LSF in Eq.~\eqref{eq:alpha:grad} is clearly not invariant under various reparameterizations and reformulations of the problem leading to the same failure set, \Fexact\, which changes the values of the LSF nonproportionally. 
In \cite{VORECHOVSKY2022115606}, binary or discrete-state functions were considered, in which case the gradient could not be used. Instead, the direction to the nearest safe point was used at every location in the failure domain; see the magenta dashed line passing through a~cloud of green dots in Fig.~\ref{fig:sensitivity}b. The density of the background in the failure domain shows that most of the contribution to $s_v^2$ is stemming from the high-density regions in $\Fexact$, and situations with multiple design points contributing to different sensitivities are treated automatically.

In this present paper, we argue that it is not necessary to base the sensitivity measures on the whole failure set \Fexact, as it is done in the case of $s_v^2$s. The sensitivities associated with individual variables be seen as contributions to the \emph{change of state} from safe operation to failure  and vice-versa. Therefore, we propose to limit the domain at which the local contributions are accounted for to the failure surface $\FS$, at which the transition between safe and failure events occurs. We expect this boundary to be sufficiently smooth. Along this boundary, the \emph{local} contributions $t_v^2(\x)$ to the transition sensitivity are again proportional to the probability density and the $\alpha$-factors, compare with Eq.~\eqref{eq:loc:contrib}
\begin{align}
 \label{eq:loc:contrib:FS}
    t_v^2(\x)
    =   f_{\X}(\x) \alpha^2_v(\x), \quad  v=1,\ldots,\Nv \, ,
\end{align}
in which $\alpha_v(\x) \equiv n_v(\x)$ are the components of the unit-length normal vector $ \pmb{n}(\x)$ at point \x\ on the failure surface. These unit normals scaled by the local density are visualized as gray arrows in Fig.~\ref{fig:sensitivity}c. The flux $T$ through the boundary is integrated over the whole failure surface to obtain the global contribution of all variables at once
\begin{align}
\label{eq:SensFluxT}
        T  = 
        \oiint\limits_{\FS}  f_{\X}(\x)  \, \dd \FS
        =
        \oiint\limits_{\FS}  \sum_{v=1}^{\Nv} 
         f_{\X}(\x) \, \alpha^2_v(\x) \, \dd \FS
        =
       \sum_{v=1}^{\Nv}  \oiint\limits_{\FS}  t_v^2(\x) \, \dd \FS \, .
\end{align}
The last expression suggests that its integral can be split into contributions of individual variables by using the squared $\alpha$ factors computed from the local geometry. Therefore, the proposed sensitivities $t_v^2$ are easily obtained as the contour integral of contributions of variable $v$ to the total flux $T$ along the whole boundary
\begin{align}
\label{eq:tv}
    t_v^2 
    = \frac{1}{T} \oiint_{\FS} t_v^2(\x)                   \, \dd \FS
    = \frac{1}{T} \oiint_{\FS} f_{\X}(\x) \,  \alpha^2_v(\x)  \, \dd \FS
    \, .
\end{align}
The sum of all sensitivities is equal to one: $ \sum_{v=1}^{\Nv} t_v^2 = 1$ and provides the global shares of individual variables on the transition between the two states. The definition is illustrated in 
Fig.~\ref{fig:sensitivity}c.  The computation is not performed in the \Nv\ dimensional volume as in the case of $s_v^2$ \cite{VORECHOVSKY2022115606}, but it is limited to the failure surface, which is an $\Nv-1$ dimensional object only. Note also that the proposed sensitivities of transition can be generalized to cases with more than just two states (failure/success). If the corresponding sets are disjoint, the integral is limited to all boundaries between a pair of states under consideration. For example, one can be interested in the transition from the safe operation to failure type 1 independently of the transition from the safe state to failure type 2.

The practical computation of $t_v^2$ factors can be obtained as a by-product of the proposed geometrical decomposition of the design domain; see  Fig.~\ref{fig:sensitivity}d. Since we are only interested in the transitions between states, it suffices to focus on the mixed simplices in which the transition supposedly occurs. Assume the current decomposition of the explored part of the design domain is known as in the case of Fig.~\ref{fig:sensitivity}d with 85 ED points. The list of all mixed simplices is labeled $j = \{ 1, 2 , \ldots, N_{\pazocal{M}} \}$. Each such simplex contributes to $p_\pazocal{M}$ by its probability content $p_j$. In every such simplex, the contributions $\alpha_v^2$ of individual variables to probability $p_j$ can be partitioned by using the approximation of the normal to the true failure surface. To obtain these normals, we propose to (1) divide all $N_{\pazocal{M}}$ simplices into subgroups which are connected via $\Nv-1$ dimensional facets (form a solid body of mixed simplices),  (2) divide each such subgroup into the largest possible clusters of simplices which can be cut by a single (hyper)plane strictly separating ED points (vertices) of different types (3) use the plane normals $\pmb{n}_{j}$ for all the simplices $j$ in the cluster.
The illustration in Fig.~\ref{fig:sensitivity}d shows a case in which the failure surface is covered by a compact set of mixed simplices, the pairs of which share a line segment (a single subgroup). However, there are a~limited number of unique directions approximating the normals obtained from the clusters of adjacent simplices. This clustering is important to avoid potential spurious sensitivities, which would be obtained if the normal directions are approximated in each simplex separately. Small deviations of these approximated normals would artificially inflate some $t_v^2$ factors at the expense of others, because the fragmented normals would have parasite projections onto some variables.
Once the probabilities $p_j$ in all simplices are split into the individual variables, the sum of these contributions is divided by the sum of probabilities to form the estimated sensitivities
\begin{align}
\label{eq:tv:est}
    t_v^2 
    \approx \frac {\sum \limits_{j \in \Mn{\Ns} }
     \left( p_j \cdot V_j^{-1/\Nv} \right)  n^2_{j,v}    }
            {\sum \limits_{j \in \Mn{\Ns} }
             \left( p_j \cdot V_j^{-1/\Nv} \right)         } 
             \, ,
\end{align}
where the term $  p_j \cdot V_j^{-1/\Nv}$
approximates the integral $ \oiint  f_{\X}(\x)  \, \dd \FS$ taken over the $j$th simplex. In this approximation, we assume that the $j$th simplex has a~regular shape, and its volume $V_j = S_j \cdot v_j $ can be written as the product of the surface area $S_j$ through which the flux takes place, and thickness $v_j$, 
    where $S_j = V_j^{(\Nv-1)/\Nv}$ and 
          $v_j = V_j^{1/\Nv}$.
The probability content of the $j$th simplex can be written as the product of its average density $ f_j$ and volume:
$  p_j = f_j \cdot V_j$. Therefore, the required average flux 
$  f_j \cdot S_j = p_j \cdot V_j^{-1/\Nv}$. 



\section{Numerical Examples with Independent Gaussian Variables 
\label{sec:2D}
}

We demonstrate the proposed method using numerical examples, starting with independent Gaussian input random variables and ending with an illustration of employing the Nataf transformation for the case of correlated non-Gaussian variables. The first eight examples feature standardized \emph{bivariate} problems with the performance functions selected to explore the problems which are posing unique challenges to the modern existing algorithms (multiple failure points, dependence on a~problem reparameterization, alternating domains, closed failure domains, etc.). The ninth example is a~practical problem featuring three Gaussian variables, and it is followed by a 6D standard numerical example. To provide a picture about the behavior of the proposed method in higher dimensions, we generalize five representative types of LSFs into an arbitrary dimension and provide results from all spaces between two dimensions to 8D, which is about the limit of the proposed approach. In all numerical examples, the full definition of LSF, which may be smooth, is reduced to just a~\emph{binary function} which provides outcomes for the proposed algorithm. We can say that, generally, all existing methods which use the numerical values of LSF to  estimate the distance from failure surface, compute LSF gradients, or build a smooth surrogate, fail in such a~ setting. We demonstrate that the proposed method provides a~fast convergence towards exact results and reusable information for further refinement.

\subsection{Four-branch Function}

\begin{figure*}[!htb]
    \centering
    \includegraphics[width=\textwidth]{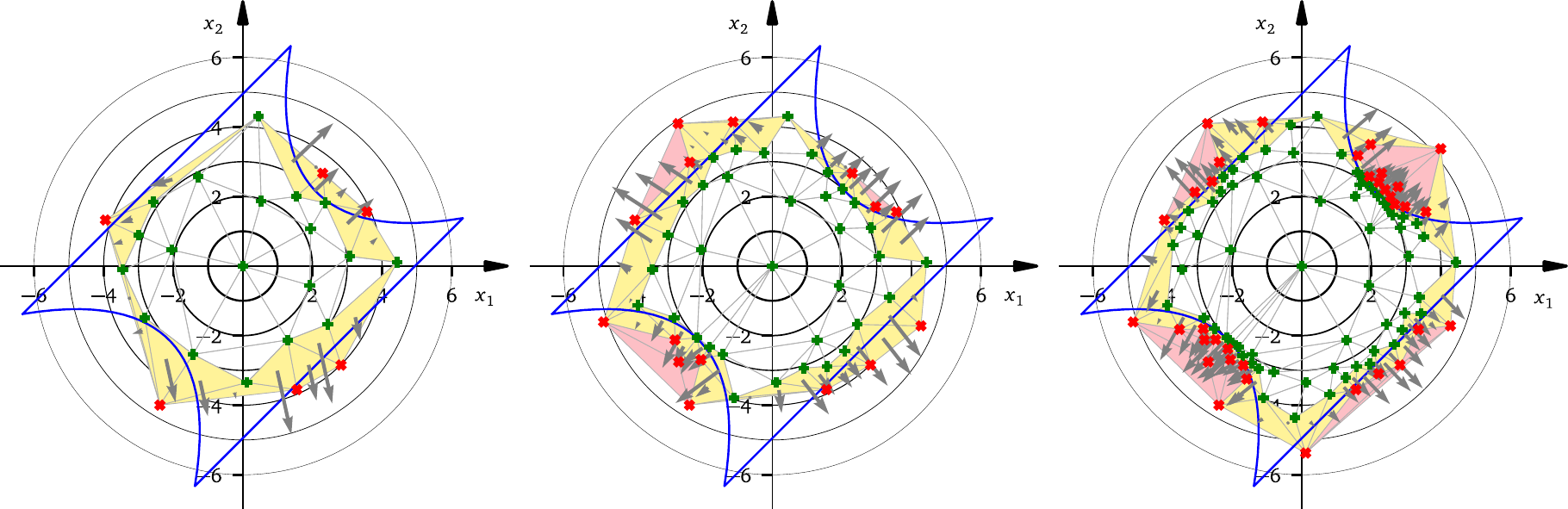}
    \caption{Four-branch reliability problem in Eq.~\eqref{eq:FourBranch}.
    Left: situation after $\Ns = 25$ model evaluations (the six detected failure points enabled the construction of 18 mixed simplices, 
    Middle: $\Ns = 50$ out of which 15 are failures (five failure simplices were constructed),
    Right:  $\Ns = 100$, the probability content outside the convex hull became orders of magnitude less than the failure probability. 
    Gray arrows show the contributions of mixed simplices to sensitivities.
    }
    \label{fig:testcases2d:fourbranch:points}
\end{figure*}

\begin{figure}[!htb]
    \centering
    \includegraphics[width=9cm]{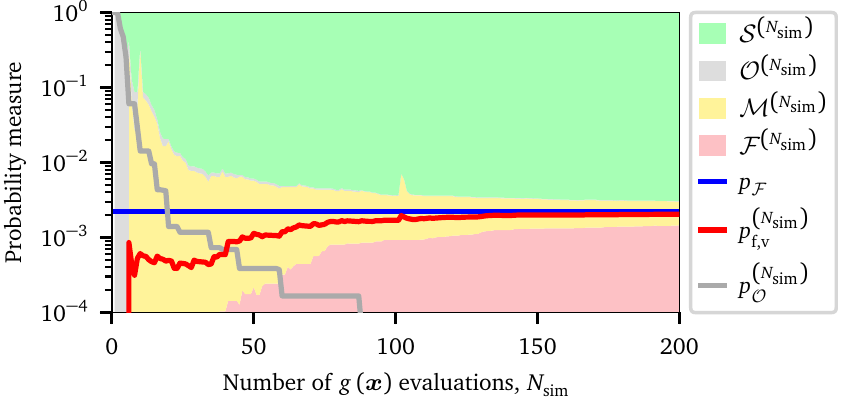} 
    \caption{ 
    Monitoring of convergence of estimations for the four-branch problem.}
 \label{fig:testcases2d:fourbranch:diagram}
\end{figure}

\begin{figure}[!htb]
    \centering
    \includegraphics[width=9cm]{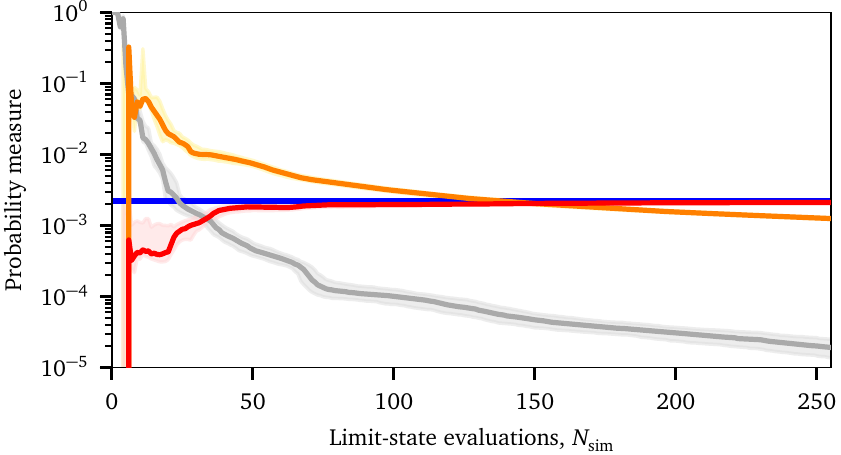} 
    \caption{Variability of the proposed method studied for the case of the four-branch example. Each median estimation is accompanied by a~band showing $25^{\mathrm{th}}$ and $75^{\mathrm{th}}$ percentiles.
    }
 \label{fig:testcases2d:fourbranch:average_diagram}
\end{figure}

The first problem is the famous four-branch function  \cite{Borri1997}. It is a~common benchmark problem in reliability analysis; see e.g. studies with various parameter settings \cite{Schueremans2005, Echard2011, Papaio:Papa:Straub:SeqIS:SS:16, Schbi2017, LIU2019102687:fourbranch, BALESDENT20131:fourbranch, JIANG201947:fourbranch, XIONG2021107693:fourbranch, BAO2021107778:fourbranch:modified, ZHANG2019440:fourbranch:modified, vahedi2018structural:fourbranch:modified, SUN2017152:fourbranch} even though the exact failure probability is typically quite high, making the problem easy to solve.
The function describes the failure of a~\emph{series system} with four distinct limit state components: two linear and two nonlinear branches of the failure surface
\begin{align}
  \label{eq:FourBranch}
    g \left( x_1, x_2 \right)
    &= \min
     \left\{\begin{array}{ll}
        3 + 0.1 \left( x_1 - x_2 \right)^2  - {\left(x_1+x_2\right)}/{\sqrt{2}} \\
        3 + 0.1 \left( x_1 - x_2 \right)^2  + {\left(x_1+x_2\right)}/{\sqrt{2}} \\
        x_1 - x_2 + 7/\sqrt{2}\\
        x_2 - x_1 + 7/\sqrt{2}
        \end{array},\right. 
\end{align}
where the two input variables $\x=\{x_1,x_2\}$ are modeled as two independent standard Gaussian random variables. Note that various authors use various parameters replacing the number $7$ (originally 3.5, sometimes 6, or 11). The failure event is defined as $g \leq 0$ and the proposed algorithm is using this binary information only.  
The blue line in Fig.~\ref{fig:testcases2d:fourbranch:points} represents the failure surface which separates the closed safe region from the open failure region. The concentric circles illustrate the contours of identical Gaussian density. 
Fig.~\ref{fig:testcases2d:fourbranch:points} also shows the constructed safe, mixed, and failure simplices at three stages of the adaptive sequential sampling: after evaluating $\gx$ 25 times, 50 times, and 100 times.
The failure probability estimation of the proposed method after $\Ns=200$ LSF evaluations is $2.03 \cdot 10^{-3}$, which is quite close to the exact solution $\pF = 2.22 \cdot 10^{-3}$.

The evolution of the proposed estimators is reported in Fig.~\ref{fig:testcases2d:fourbranch:diagram}, which will now be described as a~typical convergence plot for the proposed method.
Typically, it takes a~number of model evaluations to construct several ``green'' simplices bounding regions deemed as safe. The domain outside the regions covered by simplices is represented by gray color. It can be seen that the fifth \gx{} function evaluation leads to the discovery of the first failure event; see the first nonzero estimations of $p_{\mathrm{f,v}}$ via Eq.~\eqref{eq:vertex2}, and also the appearance of mixed simplices. Since that moment, the extension algorithm kept switching between the exploration of new outside territories and exploitation of the existing mixed regions (yellow simplices). At $\Ns= 29$ LSF evaluations, the first closed failure triangle was constructed.
At 60 evaluations, the probability corresponding to all the red simplices became greater than the ``outside probability'' corresponding to the region outside the convex hull; see the intersection of the gray line with the purple/yellow boundary in Fig.~\ref{fig:testcases2d:fourbranch:diagram}. Since that moment, more points were sampled from inside the existing mixed simplices (refinement of the failure surface) than the points leading to the expansion outside the existing convex hull.

A natural question arises about the reproducibility of the results obtained with the proposed algorithm, see  Sec.~\ref{sec:convergence}. 
The four-branch problem was run $\Nr = 500$ times with the same parameter settings, and we present the convergence of estimators in Fig.~\ref{fig:testcases2d:fourbranch:average_diagram}. The thick solid lines represent the medians, and they are each accompanied with a~scatter-band showing the 25$^{\mathrm{th}}$ and 75$^{\mathrm{th}}$ percentile. The random scatter is very low, i.e., the scaffold of simplices is qualitatively similar in various runs -- it behaves almost like a~deterministic method. The only way to change the character of these curves is to modify the aggressiveness of the exploration; see the discussion in Sec.~\ref{sec:extension}. 
Apart from the estimators presented previously in Fig.~\ref{fig:testcases2d:fourbranch:diagram}, we added an orange line showing the probability associated with ``mixed'' simplices. When this line crosses the line corresponding to the estimated failure probability $(\Ns \approx 150)$, the probability content associated with simplices containing the boundary becomes so low that its further refinement may be found unnecessary.

\paragraph{Comparison with Other Existing Methods in the Literature}

\begin{figure*}[!htb]
    \centering
    \includegraphics[width=\textwidth]{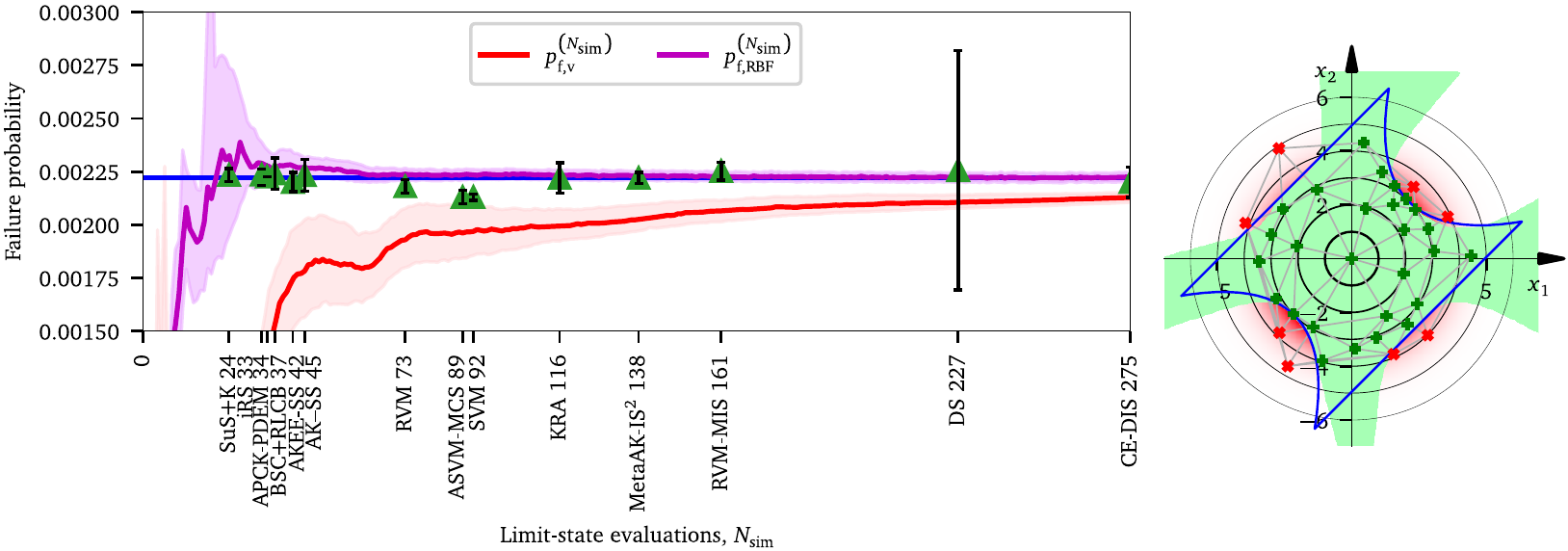} 
    \caption{Left: comparison with other methods. the whiskers delimit the interval $\pm0.67 \sigma$, 
    strip bands show  $25^{\mathrm{th}}$ and $75^{\mathrm{th}}$ percentiles. 
    Right: radial basis function surrogate model based on ED after $\Ns=35$ LSF evaluations.
    }
 \label{fig:testcases2d:fourbranch:comparison}
\end{figure*}

We use this example to also compare the efficiency of the proposed methods for the \emph{extension} of ED and the \emph{estimation} of failure probability with other existing methods.
This numerical example was studied by many authors in the past, and we collected the results from the literature in Tab.~\ref{tab:comparison}. The same data is also plotted in Fig.~\ref{fig:testcases2d:fourbranch:comparison}, in which the average results are accompanied by estimations of variance. 
\begin{table}[!ht]
\label{tab:comparison}
\caption{Summary of papers where the four-branch problem has been studied.}
\centering
\begin{tabular}{lllll}
  \toprule
  Year & Method & $\Ns$ & $\hat{\pf} \cdot 10^3$ & $\mathrm{cov}_{\hat{\pf}} \left(\%\right)$ \\ 
  \midrule
  
    
    
2000 & DS \cite{Waarts2000:fourbranch} & 227 & $2.2556$ & 37\\
    
    
2011 & AK-MCS+U \cite{Echard2011} & 96 & $2.233$ & -\\
2011 & AK-MCS+EFF \cite{Echard2011} & 101 & $2.232$ & -\\

2011 & ${}^2$SMART \cite{BOURINET2011343:fourbranch} & 1035 & $2.21$ & 1.7\\
    
    
2013 & CE-AIS-GM \cite{KURTZ201335:fourbranch} & 3943 & $2.15$ & 3\\

2014 & MetaAK-IS$^2$ \cite{CADINI2014109:fourbranch} & 48+90 & $2.22$ & 1.7\\
    
2016 & AK–SS \cite{HUANG201686:fourbranch} & 45 & $2.233$ & 4.94\\
    
2017 & KRA \cite{XUE20171:fourbranch} & 116 & $2.22$ & 4.7\\

2017 & ASVM-MCS \cite{Pan2017} & 89 & $2.13$ & 2.2\\

2018 & iRS \cite{GUIMARAES201812:fourbranch} & 33 & $2.24$ & -\\


2019 & AKEE-SS \cite{ZHANG201990:fourbranch} & 41.7 & $2.20$ & 3.07\\
    
    
2020 & BSC+RLCB \cite{yi2020efficient:fourbranch} & 36.66 & $2.240$ & 4.89\\

2020 & DRL \cite{XIANG2020106901:fourbranch} & 2597 & $2.314$ & -\\
    
2021 & ABSVR1 \cite{WANG2021114172:fourbranch} & 39.6 & $2.214$ & -\\
2021 & ABSVR2 \cite{WANG2021114172:fourbranch} & 42.8 & $2.222$ & -\\
    
2021 & SuS+K \cite{RePEc:fourbranch} & 23.9 & $2.234$ & 2.05\\

2021 & RVM \cite{LI2021381:fourbranch} & 73 & $2.18$ & 2.14\\
    
2021 & SVM \cite{LEE2021107481:fourbranch} & 91.97 & $2.13$ & 10\\
    

2022 & RVM-MIS \cite{WANG2022108287:fourbranch} & 161 & $2.252$ & 2.759\\
    
2022 & CE-DIS \cite{ZHANG2022108306:fourbranch} & 275 & $2.20$ & 4.79\\
    
2022 & APCK-PDEM \cite{ZHOU2022108283:fourbranch} & 34.5 & $2.226$ & 0.32\\

\bottomrule
\end{tabular} 
\end{table}

The red line shows the results obtained with the proposed method in the basic setting: the estimators are based on just binary information from the limit-state function. Therefore, the accuracy of the estimations are worse than for the other methods under comparison which are building various surrogate models and exploit the information about numerical values of the smooth $\gx$ function.

When the numerical values produced by the \gx\ function are trusted to map a~well-behaved, continuous, and smooth output, the values in the vertices can be used to construct a~surrogate model (one global approximation or a~number of local interpolations). Such surrogate has the potential to approximate the true failure surface very accurately as it is based on a~collection of ED points placed close to the failure surfaces in high probability regions (see the proposed active learning in Sec.~\ref{sec:extension}). These surrogate models can be formed by, e.g., regression with polynomials, radial basis functions, Kriging, Polynomial Chaos Expansion, Neural Network, etc. The obtained surrogate function can then be analyzed, e.g., via a~sampling estimation such as Importance Sampling. In order to get a~fair comparison of the proposed \emph{active learning} \& \emph{rare-event probability estimation} with other methods that are using the numerical value of the LSF, we performed another set of computations. We reused the ED points previously obtained from the ED \emph{extension}, but for each sample size, we set up a~radial basis function (RBF) approximation based on the $\gx$ function values, not just the binary outcomes. We used the simplest RBF with the default parameters in class \lstin{Rbf} from the \lstin{scipy.interpolate} sub-package \cite{2020SciPy-NMeth} to perform the classification into 
failure and safe domains based on the sign of the RBF surrogate. 
The results of repeated runs are presented in Fig.~\ref{fig:testcases2d:fourbranch:comparison} left using the magenta color (median $\pm$ $25^{\mathrm{th}}$ and $75^{\mathrm{th}}$ percentiles). It is no surprise that the representation of the simple LSF function is excellent for ED as small as $\Ns=35$ points, see Fig.~\ref{fig:testcases2d:fourbranch:comparison} right. 
The improvement in the estimations (Fig.~\ref{fig:testcases2d:fourbranch:comparison} left) is due to the more accurate classification of the failure surface in the vicinity of the four design points. Simply put, the ED points selected by the proposed active learning are also well suited for smooth surrogate models, despite the fact that the selection criterion itself does not use the numerical values of the \gx\ function.
It is true that the ``corners'' were not classified correctly either by RBF or by the binary surrogate, see Fig.~\ref{fig:testcases2d:fourbranch:comparison} right. However, these remote regions are irrelevant for failure probability estimation, and the extension algorithm correctly ignored these regions and focused the ED points around the boundary in high probability regions. 
The efficiency of the RBF classifier is as good as the efficiency of the best methods known from the literature. It means that the proposed method is competitive for well-behaved LSFs, but its strengths stand out for harder problems.
Our experience with the other methods which are building surrogate representations based on Kriging or Support vector regression is that they depend on the user through the selection of a~kernel and its parameters too heavily.
Indeed, authors of some papers even state that for this example, they made an exception and skipped the screening  or ``intuitively'' set up regions of interest by hand. We consider this to be impractical for wide applicability of algorithms in practice. We declare that all numerical examples in this paper are computed with the very same settings of the extension algorithm. 
In the proposed method, there is no need for the user to intervene.

The proposed global sensitivities are identical $ t_{1}^2  = t_{2}^2  =  0.5$ due to symmetry reasons.








\subsection{Piecewise Linear Functions
\label{sec:example:piecewise}}

Consider a~series system involving two limit state functions involving again two independent standard normal variables $x_1$ and $x_2$:
$g(\x) = \min \left(  g_1 , g_2 \right)$. These two limit state functions are simple piecewise linear functions
\begin{align}
 \label{eq:PWL:1}
    g_1 \left( x_1 \right)
    &=
     \left\{\begin{array}{ll}
        0.85 - x_1/10, & \text{for } x_1 \leq 3.5 \\
        4 - x_1, & \text{for } x_1 > 3.5 
        \end{array}\right. 
        \, ,
    \\
        \label{eq:PWL:2}
     g_2 \left( x_2 \right)
    &=
     \left\{\begin{array}{ll}
        2.3 - x_2, & \text{for } x_2 \leq 2 \\
        0.50 - x_2/10, & \text{for } x_2 > 2
        \end{array}\right. 
        \, .
\end{align}
This example with two ``design points'' was presented by \citet{Breitung:19:RESS}, who demonstrated that a~simple gradient search algorithm and also the stochastic gradient algorithm (SuS) initiated from around the origin will be lead away from the most central failure point and erroneously discover only the second (more distant) local point.

\begin{figure}[!htb]
    \centering
     \includegraphics[width=9cm]{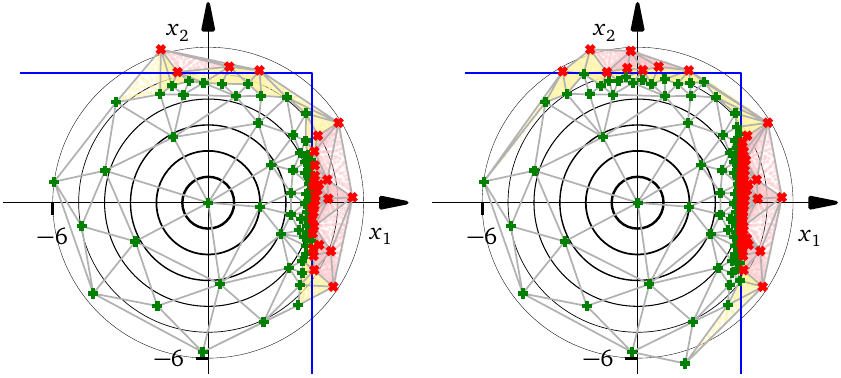}
    \includegraphics[width=9cm]{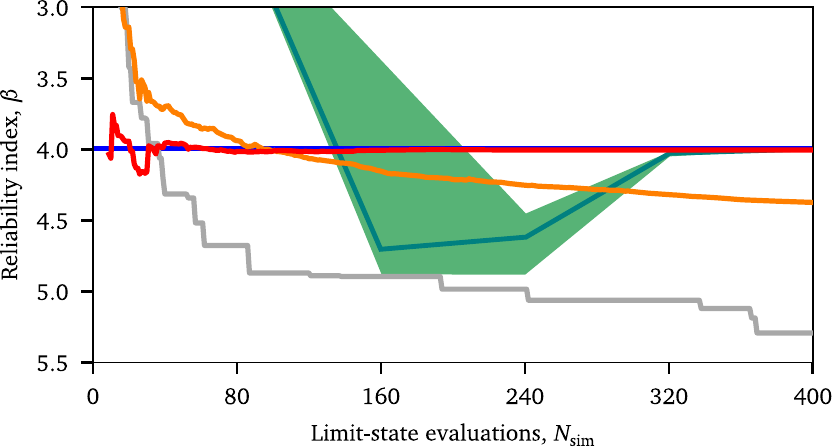}
    \caption{Piecewise linear function featuring the minimum of Eqs.~\eqref{eq:PWL:1} and \eqref{eq:PWL:2}. Plots for $\Ns = 100$ and $200$ (top) and a~convergence diagram (bottom). 
    The purple and yellow points are the deterministic quadrature points in failure and undecided simplices respectively. They are used to integrate their probability contents.
    The green line with the scatter band is redrawn from~\cite{WAGNER2022102179:piecewise}. it represents a~single run of the algorithm and the estimated $95\%$ confidence interval. The remaining lines were obtained with the proposed algorithm, and they can be interpreted using the key in Fig.~\ref{fig:testcases2d:fourbranch:diagram}.
    }
 \label{fig:testcases2d:piecewise_linear}
\end{figure}

Unlike SuS, our algorithm expands the convex hull occupying the space uniformly from the origin and cannot be misled by the values of the \gx\ function, because it only gets binary information. Fig.~\ref{fig:testcases2d:piecewise_linear} presents two stages of the ED, and also the convergence plot obtained with the proposed algorithm.
In order to compare our algorithm with a~very recent study~\cite{WAGNER2022102179:piecewise}, convergence is shown against the reliability index $\beta$ (see the green line and a~scatterband in~Fig.~\ref{fig:testcases2d:piecewise_linear}). The proposed method provides more accurate results with less LSF evaluations, and it resolves the more central part of the failure surface with more ED points -- proportionally to the probability content.


The proposed global sensitivities are
 $ t_{1}^2 
    = \varphi(4)\Phi(5) /T 
    \approx  0.989$ 
 and 
$ t_{2}^2 
    = \varphi(5)\Phi(4) /T  
    \approx  0.011$, 
where $T= \varphi(5)\Phi(4) + \varphi(4)\Phi(5) \approx \varphi(5) + \varphi(4)$,  $\varphi(x)$ is the density and $\Phi(x)$ is the cumulative distribution function of a~standard  Gaussian variable. The proposed algorithm approximates these exact values using mixed simplices. 
After the LSF evaluation in $\Ns=400$ ED points, the two sensitivities stabilize at the exact values.

%

\subsection{Invariance (a~Series System)}


In his Sec.~5.3, \citet{Breitung:19:RESS} presented another example, using which he documented the issues in SuS which are  related to gradient optimization using the shape of LSF.
He considered a~series system consisting of two independent components represented by standard normal variables $x_1$ and $x_2$. A~failure event occurs when at least one of the two components fails. In that example, the failure of the first component occurs if $x_1 >5$ and the failure of the second component is triggered when $x_2 < -4$. These conditions for individual components can be formulated, for example, using two linear functions as
\begin{align}
    g_1 \left( x_1 \right)
    &= 5 - x_1
    \\
    \label{eq:invariance:g2:lin}
     g_2 \left( x_2 \right)
    &= 4 + x_2 
    .
\end{align}
The LSF of the whole series system then reads
\begin{align}
    \label{eq:LSF:Invariance}
    g(x_1, x_2) = \min \left[  g_1(x_1) , g_2(x_2) \right]
    .
\end{align}

The same failure set can, however, be obtained by reformulating the failure condition for the second component. Breitung \cite{Breitung:19:RESS} selected a~logistic function 
\begin{align}
    \label{eq:invariance:g2:log}
     g_2^{\star} \left( x_2 \right)
    &= \frac{1}{1+ \exp \left( -2x_2-8\right) } - 0.5
\end{align}
to replace $g_2(x_2)$ in the LSF in Eq.~\eqref{eq:LSF:Invariance}. The important difference between the definitions in  Eqs.~\eqref{eq:invariance:g2:lin} and \eqref{eq:invariance:g2:log} is the shape of the LSF in the neighborhood of the origin. 
His point was to document that the same failure set expressed though a~different \emph{smooth} LSF may affect the result of a~gradient-based algorithm involved in the design point search and, similarly, it may also negatively influence the behavior of the SuS algorithm. 
Again, the problem features a~convex polyhedral safe set bounded by two orthogonal linear safety margins with reliability indices $\bm{\beta} = \{4 ; 5 \}$. Therefore, it has the same topology as in the previous example of piecewise linear functions shown in Fig~\ref{fig:testcases2d:piecewise_linear}. 
The failure probabilities corresponding to the two failure surfaces independently are of different orders of magnitude: 
$\pF (\beta_1) = \Phi(-4) \approx 3.167\cdot10^{-5}$ and
$\pF (\beta_2) = \Phi(-5) \approx 2.867\cdot10^{-7}$. The probability of failing both criteria  simultaneously is negligible
$\pF = \pF (\beta_1) + \pF (\beta_2)  - \left[ \pF (\beta_1) \pF (\beta_2)  \right] \approx 3.196\cdot10^{-5}$.
The proposed global sensitivities are swapped from the previous example, i.e., 
$ t_{1}^2  \approx  0.01 $ 
 and 
$ t_{2}^2   \approx  0.989$.



This dependence on the various definitions of the same failure condition, and the related convergence problems that can occur for more complex systems, supports our argument for having the algorithm based on a~dichotomous definition: failure or success. The proposed algorithm is not affected by reformulation or reparameterization of the underlying reliability problem, and the running estimations quickly converge towards the exact solution.

\subsection{Product Function}

\begin{figure}[!htb]
    \centering
    \includegraphics[width=9cm]{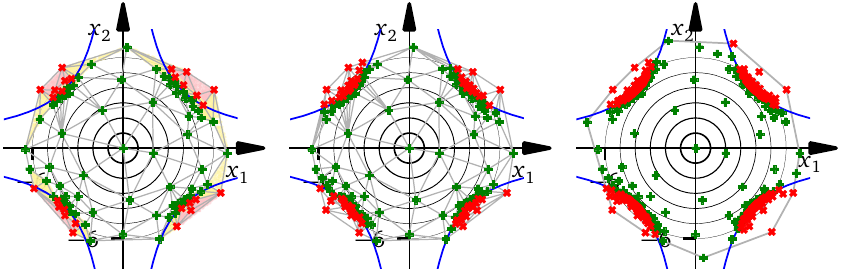}
    \includegraphics[width=9cm]{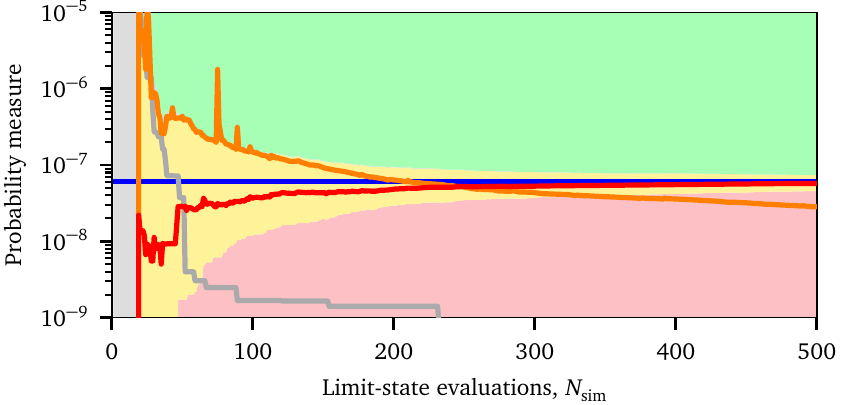} 
    \caption{The product function defined in Eq.~\eqref{eq:product:g}  with four design points. 
    Top left: situation after $\Ns = 100$ model evaluations, with integration nodes in failure and mixed simplices,
    top middle: $\Ns = 200$,
    top right:  $\Ns = 500$, only convex hull is displayed.
    Bottom: Convergence of the estimates using the key in Fig.~\ref{fig:testcases2d:fourbranch:diagram}.
    }
 \label{fig:testcases2d:prod4betas}
\end{figure}

This example illustrates the problem for which a~gradient-based method  ``slides down'' in the direction of the steepest descent, but it is generally unable to find \emph{all} design points. 
Let us define a~``product'' function which has four design ``beta'' points,
\begin{align}
 \label{eq:product:g}
     g\left( x_1, x_2 \right)
    &= \beta^2/2 - \left| x_1 \cdot x_2 \right|
    ,
\end{align}
in which the two input variables $X_1, X_2$ are independent standard normal variables, and negative values of \gx\ signal failure.
\citet{Breitung:19:RESS} set  $\beta=\sqrt{30}$ and reported that in his simulations SuS found all four beta points only in nine out of fifty runs of the method. The performance of the proposed algorithm is reported in Fig.~\ref{fig:testcases2d:prod4betas}. 
The algorithm expands uniformly in all directions without prejudices and reliably finds all four design points. When the number of function calls exceeds $\Ns=100$, the outside probability drops below the estimate failure probability and the parts of the boundary separating all four important failure regions are well refined.
The analytical solution can be based on the fact that the product $Z= X_1 X_2 $ has the probability density $f_Z \left(z\right) = K_0\left(|z|\right)/\pi$, where $K_n\left(\cdot \right)$ is the modified Bessel function of the second kind ($n$th order). The distribution function
$F_Z \left(z\right) = \int_{-\infty}^z f_Z \left(t\right) \mathrm{d}t $ can be solved in a~closed form as follows
\begin{align}
    F_Z \left( z \right)= 
    \frac{1}{2}+
    \frac{z}{2\pi}
    \Big[
    K_0\left( \vert z \vert  \right) \left(2+\pi L_1\left(z \right) \right) 
    +
    K_1\left( \vert z \vert \right)  \pi L_0\left( \vert z \vert \right)
\Big] \, ,
\end{align}
where $L_0\left(\cdot \right)$ and $L_1\left(\cdot \right)$ are the modified Struve functions of orders zero and one, respectively. 
Using the distribution function, the exact failure probability reads:
$
    \pF
  = P\left(\gX < 0 \right)
  = P\left(\beta^2/2 < \left| X_1 \cdot X_2 \right| \right) 
  = P\left(Z < -\beta^2/2 \; \cup \; Z > \beta^2/2 \right)
  = 2 \cdot F_Z \left( -\beta^2/2 \right)
\approx 6.064 \cdot 10^{-8}$. Obtaining an accurate estimation of such a~small failure probability in the presence of multiple design points -- binary LSF with a~nonlinear failure surface with only about one hundred LSF evaluations -- is quite good. 

The proposed global sensitivities quickly converge to identical values $ t_{1}^2  = t_{2}^2  =  0.5$ due to symmetry reasons.


\subsection{Metaball Function}

\begin{figure}[!htb]
    \centering
    \includegraphics[width=9cm]{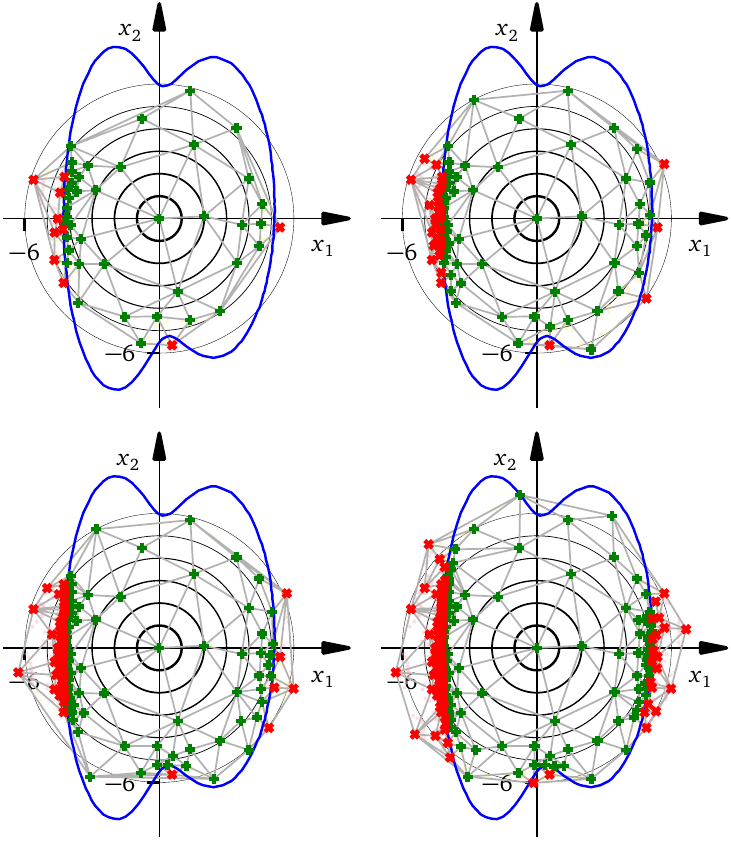}
    \includegraphics[width=9cm]{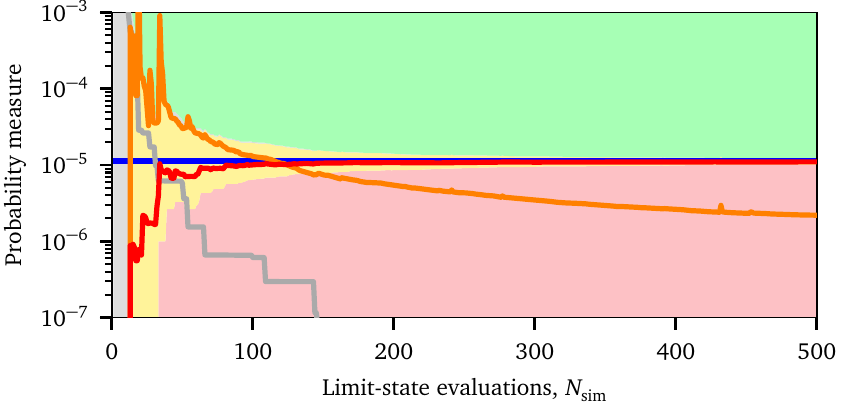} 
    \caption{Metaball function defined in Eq.~\eqref{eq:metaball}. Top: $\Ns = 50, 100, 200$ and $500$ limit-state evaluations. Bottom: Convergence of estimations.}
 \label{fig:testcases2d:metaball}
\end{figure}

\citet{Breitung:19:RESS}, inspired by \cite{wiki:Metaballs}, used the Metaball example to document the risk that SuS does not move its stochastic gradient optimization towards the region with the highest contribution to failure probability and becomes sidetracked towards far less probable failure regions. The particular definition of the Metaball function in \cite{Breitung:19:RESS} reads
\begin{align}
\label{eq:metaball}
\begin{split}
     g\left( x_1, x_2 \right)
    &= \frac{30}{
        \left( 
            \frac{4\left( x_1 + 2\right)^2 }{9} +
            \frac{x_2^2}{25}
        \right)^2 + 1 
    }\\ &+ 
    \frac{20}{
        \left( 
            \frac{\left( x_1 - 2.5\right)^2 }{4} +
            \frac{\left( x_2 - 0.5\right)^2 }{25}
        \right)^2 + 1 
    } - 5
    .
\end{split}
\end{align}
It is a~hard problem that misleads the downhill-based methods with its complicated geometry. The function has a~saddle point located at the origin (mean values). Downhill optimization leads to quite distant failure regions while the most central failure point is behind a~high ``hill''.  The topological change  confuses some algorithms, because their presumed representation of failure surface topology is wrong. The proposed algorithm is absolutely robust as it does not use the function value to orient the exploration and keeps refining the four consecutively discovered failure regions proportionally to their probability contributions in a~balanced manner; see Fig.~\ref{fig:testcases2d:metaball}.
Occupying the unexplored domains while refining the existing boundaries proportionally to the probability gains warrants that 
  (i) the regions with a~large probability content will be covered by points and 
  (ii) the yet undiscovered failure modes will be revealed.
A detailed analysis yields $\pF \approx 1.12857 \cdot 10^{-5}$. 
The point estimator of $\pF$ after 100 and 500 LSF evaluations in the proposed algorithm were $
(\cdot 10^{-5})$:         $\pfv[100]=1.007$ and 
 $\pfv[500]=1.101 $.


The excellent accuracy of these results contrasts with the quite inaccurate estimates obtained in the recent paper \cite{AMERYAN2022108036:metaball:suspension3d} using the 
AK-SESC procedure: $p_{\mathrm{f,AKSESC}}^{\left(5954\right)} = 1.59 \cdot 10^{-5}$ after $\Ns=5954$ evaluations.

The proposed global sensitivities are 
$ t_{1}^2 \approx 0.987 $ 
and 
$ t_{2}^2 \approx 0.013$.


\subsection{Narrow V-shaped Failure Band -- Late Discovery \& Refinement}



    
        


\begin{figure}[!htb]
    \centering
    \includegraphics[width=9cm]{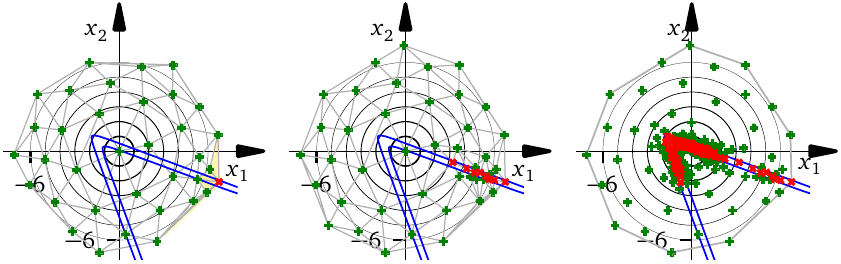} 
    \includegraphics[width=9cm]{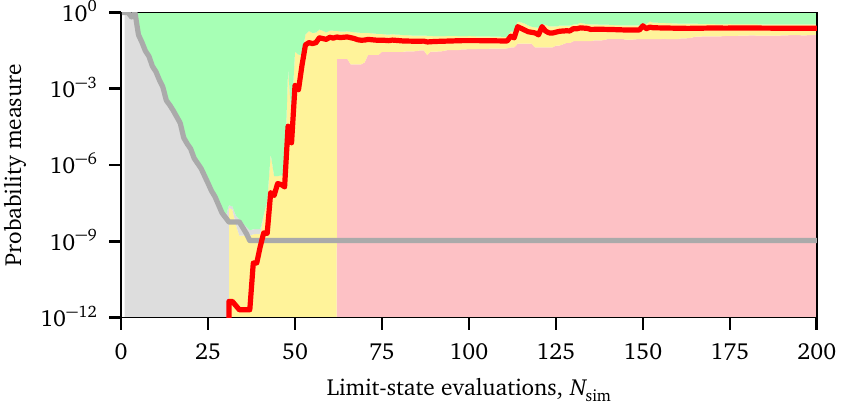} 
    \caption{
    The narrow failure domain problem in Eq.~\eqref{eq:fajfka}.
    Top left: situation after $\Ns = 35$ model evaluations, with integration nodes in failure and mixed simplices,
    top middle: $\Ns = 50$,
    top right:  $\Ns = 200$, only convex hull plotted.
    Bottom: estimation graph.
    }
 \label{fig:testcases2d:five}
\end{figure}

Consider the following limit state function described using polar coordinates centered at point $\left(-1.8, 1\right)$
\begin{align}
\label{eq:fajfka}
     g\left( \rho, \phi \right)
    &=
    \rho \left|10 - 11\cos\left(\phi - \frac{\pi}{4}\right)\right| - 1.
\end{align}
The failure region in which $g<0$ is a~narrow strip between two V-shaped boundaries; see Fig.~\ref{fig:testcases2d:five}. It takes many function evaluations to hit this narrow failure region. This failure domain protrudes through the regions of high density and, therefore, the failure probability is relatively high. It is easily possible that the high-probability failure locations get initially overlooked as they get surrounded by ``safe'' ED points. These green points support the construction of a~coarse scaffold made of presumably ``safe'' simplices. Fortunately, the failure region is open, and thus it is a~matter of time when the region will be hit by an exploration point. Indeed, Fig.~\ref{fig:testcases2d:five} visualizes a~run in which the thirty-one limit-state evaluation discovered the first (and quite distant)  failure event. Since that moment, the algorithm naturally focuses on refining the boundary and backtracks the failure surfaces towards the high-probability regions. When the estimation of failure probability becomes equal to the probability corresponding to the outside (unexplored) region (at about $\Ns \approx 40$), the algorithm automatically stops expanding the convex hull towards infinity (the gray line stays horizontal) and the exploitation of the local information about mixed regions is preferred until the boundary domain is refined. After the discovery of the first failure, the emergence of additional ``red'' points resembles the flow of points through a thin tube, see the top row of Fig.~\ref{fig:testcases2d:five}.
    


The proposed global sensitivities are 
$ t_{1}^2 \approx 0.266 $ 
and 
$ t_{2}^2  \approx  0.734$.
The estimated values based on mixed simplices built after only 200 LSF calls were $\{ 0.33, 0.67\}$.

\subsection{Closed Failure Domain}

\begin{figure}[!htb]
    \centering
    \includegraphics[width=9cm]{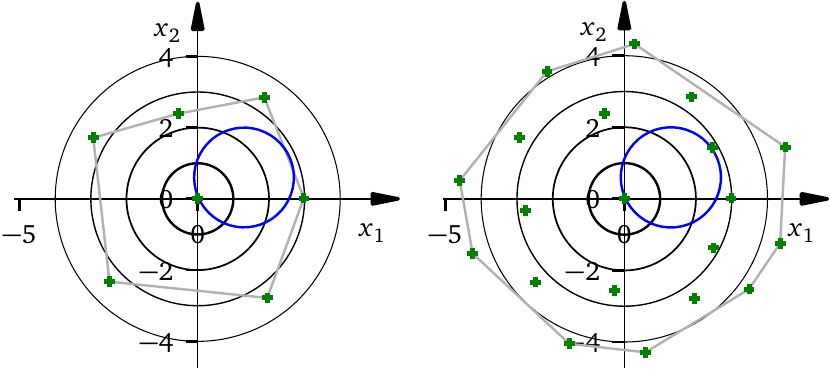} 
    \caption{Failure of the method for a~closed unsafe region. Left: situation after $\Ns = 7$ limit-state evaluations; Right $\Ns = 20$.}
 \label{fig:testcases2d:circle}
\end{figure}

We postulated that a~necessary condition for the exploratory phase to eventually hit a~failure domain is that the failure domain is open; see the previous numerical example. Indeed, closed regions are not guaranteed to be discovered. Fig.~\ref{fig:testcases2d:circle} illustrates a~failure region purposely constructed to break the method down. The closed failure domain is formed by a~circle placed at a~region with a~very high probability, but its location is so unfortunate that the exploratory phase can miss it and encapsulate it inside ``safe'' simplices. A~remedy would be the use of additional exploratory points deliberately sampled from the safe simplices to increase the chance of hitting the circle interior. Once at least one point from the unsafe region is discovered, the method automatically refines the failure surface via exploitation, as in the previous example. The next numerical example is focused on this remedy.


\subsection{Modified Rastrigin}

\begin{figure}[!htb]
    \centering
    \includegraphics[width=9cm]{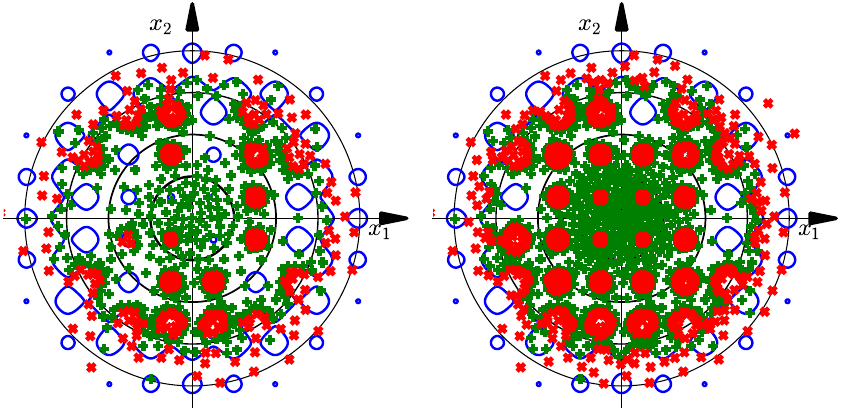} 
    \includegraphics[width=9cm]{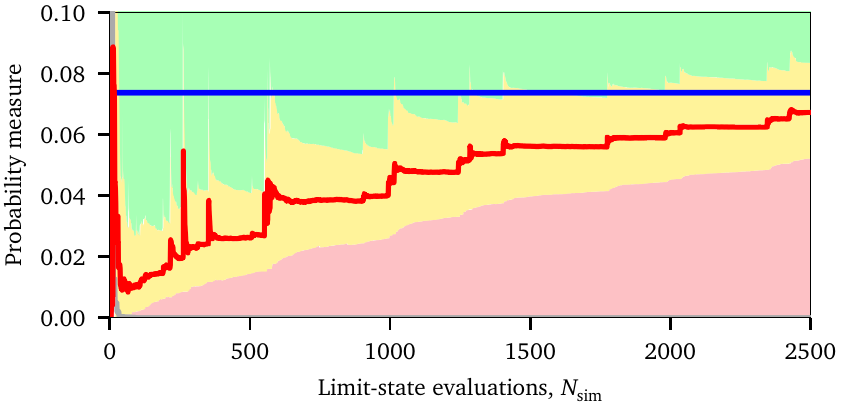} 
    \caption{Performance of the method for closed scattered regions. Top left: situation after $\Ns = 1000$ LSF  evaluations; Top right: $\Ns = 2500$.}
 \label{fig:testcases2d:rastrigin}
\end{figure}

The following example with strongly nonlinear LSF leading to many scattered failure regions was studied in~\cite{Echard2011,VORECHOVSKY2022115606}
\begin{equation}
     \label{eq:testcases_2D:rastrigin}
     g\left( x_1, x_2 \right) =
    10 - \sum_{i=1}^{2} \left(x_i^2 - 5\cos \left(2 \pi x_i\right)\right)
    .
\end{equation}
There are twenty four closed failure areas within the safe area, beyond which there is one open failure area interspersed with many closed safe areas.
Small, closed circular failure domains will almost certainly not be hit by the proposed algorithm. In order to overcome this difficulty and increase the chance of discovering closed domains, it is necessary to leave a part of the budget of points for screening, which, however, can benefit from the constructed triangulation. In this example, we let every tenth added point be selected using the global space-filling criterion, i.e., from any simplex regardless of its classification (fail/safe/mixed). 
The result of the simulation can be seen in Fig.~\ref{fig:testcases2d:rastrigin}, in which we document that many closed failure regions were initially overlooked during the expansion of the convex hull and refinement of the previously discovered mixed simplices. It took a~quite large number of LSF evaluations to identify most of the closed regions: only after ca $\Ns = 2500$ all the sixteen most central failure ``bubbles'' were hit. This translates to 250 screening samples. As visible in Fig.~\ref{fig:testcases2d:rastrigin}, the screening strategy was not yet entirely successful in this example with  high failure probability $\pF = 7.34 \cdot 10^{-2}$. The closed scattered failure regions present a~challenge to the proposed algorithm. 
The proposed global sensitivities are correctly identified as identical:
$ t_{1}^2 = t_{2}^2 = 0.5$.



\subsection{Vehicle Suspension: 3D Engineering Example
    \label{sec:suspension_3D}
}

\begin{figure*}[!ht]
    \centering
    \includegraphics[width=\textwidth]{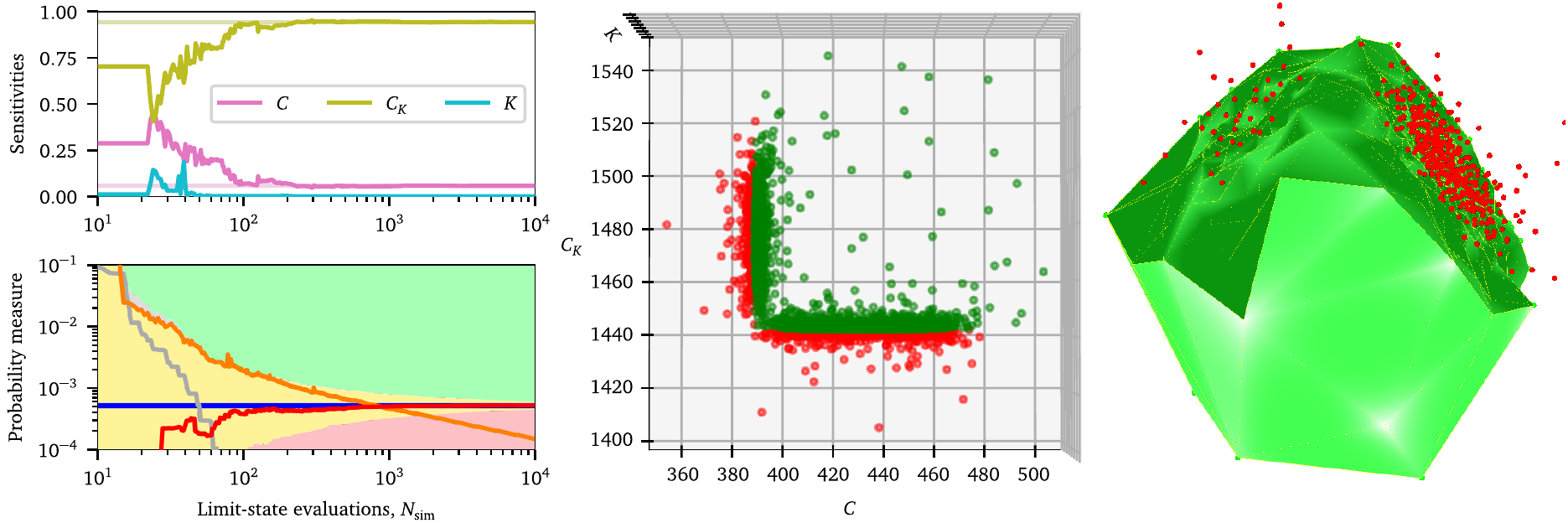}
    \caption{3D vehicle suspension example in Eq.~\eqref{eq:suspension}. 
        Left: sensitivity and probability estimations (log-log) convergence diagrams. 
        Middle: the state after $\Ns=10000$ evaluations is shown.
        Right: colored simplices for $\Ns=1\,000$. Light green facets are formed by triangles with safe points only. Dark green facets are in contact with mixed simplices.
        }
 \label{fig:suspension_3D}
\end{figure*}

\begin{table}
\centering
\caption{Input variables (both deterministic and normally distributed) for the vehicle suspension example in Sec.~\ref{sec:suspension_3D}.}
\label{tab:suspension_3D}
\begin{tabular}{llcccc}
  \toprule
  & Variable & Unit  & $\mu$ & $\sigma$ \\
  \midrule
  $A$ & road coefficient & $\mathrm{rad} \cdot  \mathrm{cm}^2  / \mathrm{m}$ & 0.15915 & -- \\ 
  $b_0$ & load coefficient & -- & 0.27 & -- \\
  $V$ &  vehicle velocity & $\mathrm{m}/ \mathrm{s}$  & 10 & -- \\ 
  $M$ &  sprung mass & $\mathrm{kg} / \mathrm{cm} / \mathrm{s}^{2}$ & 3.2633 & -- \\ 
  $m$ &  unsprung mass & $\mathrm{kg} / \mathrm{cm} / \mathrm{s}^{2}$ & 0.8158 & -- \\ 
  $g$ & gravity acceleration & $\mathrm{cm} / \mathrm{s}^{2}$ & 981 & -- \\ 
  $C$ & suspension stiffness & $\mathrm{kg} / \mathrm{cm}$ & 431.7221 & 10 \\ 
  $C_K$ &  tire stiffness & $\mathrm{kg} / \mathrm{cm}$  & 1475.5503 & 10 \\ 
  $K$ &  damping coefficient & $\mathrm{kg} /\mathrm{cm} / \mathrm{s}$ & 55.0406 & 10 \\ 
\end{tabular} 
\end{table}

This illustrative 3D example is inspired by a~design optimization of a~real engineering problem, namely, a~passive vehicle suspension~\cite{doi:10.1504/IJVD.1984.061093:suspension_3D}. The objective is to minimize the vertical vibration acceleration of the vehicle body. A~recent study~\cite{RASHKI201424:suspension_3D} generalizes the problem and solves it in the context of a~reliability-based design optimization: an optimal design should satisfy physical constrains at a~target failure probability level. 
There are various physical constrains expressing the conditions of the undesirable state occurrence. They are formulated in terms of four limit-state functions corresponding to:
    exceedance of the road-holding ability of the vehicle ($g_1$), 
    exceedance of the rolling angle ($g_2$), 
    bumper hitting ($g_3$), and 
    exceedance of the minimum required tire life ($g_4$) 
%
%
%
%
\begin{align}
    \nonumber
    g_1 \left( \x \right)
    &= 1 - \frac{\pi m V A}{b_0 K g^2} 
    \left[\left(\frac{C_K}{m + M} - \frac{C}{M}\right)^2 + \frac{C^2}{mM} + \frac{C_K K^2}{m M^2} \right]
    \\
    \nonumber
     g_2 \left( \x  \right)
    &= 4000 \cdot C \left(Mg \right)^{-1.5} - 8.6394
    \\
    \label{eq:suspension_3D:constrains}
    g_3 \left( \x  \right)
    &= 2 \sqrt{Mg \left( \frac{K^2 C_K}{C \left(m + M \right)} + C \right)} - 1 
    \\
    \nonumber
    g_4 \left( \x  \right)
    &= C_K - \left[g \left(M + m \right) \right]^{0.877}
    \, .
\end{align}
The physical meaning and numerical values of either deterministic or Gaussian variables featured in the four limit state functions are listed in Tab.~\ref{tab:suspension_3D}. Since all constrains must be satisfied simultaneously, the vehicle suspension forms a~series system, and the resulting LSF of the whole system can then be written as 
\begin{align}
    \label{eq:suspension}
    \gx = 
    \min 
        \left[  
            g_1\left( \x \right) , 
            g_2\left( \x \right), 
            g_3\left( \x \right), 
            g_4\left( \x \right) 
        \right].
\end{align}
We employed $10^6$ LSF evaluations using the importance sampling method to estimate the exact failure probability accurately: $\pF \approx 5.2 \cdot 10^{-4}$. 
The proposed algorithm delivered the following point estimations $(\cdot 10^{-4})$:
    $\pfv[100]  = 3.76$,     
    $\pfv[1000] = 4.91$, and 
    $\pfv[10000]= 5.16$; 
see~Fig.~\ref{fig:suspension_3D}. 
It is clear that as the number of points increases, the probability gain of every next point decreases. 
Indeed, when $\Ns$ exceeds approximately 100 simulations in this example, the algorithm shows exponential (power-law) convergence, which is documented by the fact that the probability corresponding to the mixed simplices forms a~straight line 
in the log-log diagram in~Fig.~\ref{fig:suspension_3D}. 

The constructed geometrical representation provides additional valuable insights into the problem. It reveals that there are actually two dominant failure regions corresponding to minimum levels of suspension $C$ and tire stiffnesses $C_K$, i.e., variables marked as $x_1$ and $x_2$ in Fig.~\ref{fig:suspension_3D}. Indeed, since the analytical expression for the individual functions $g_i$ are available in this example, we confirm that failure occurs
 when the suspension stiffness $C \leq 391.212$ 
     (negative $g_2$) or 
 when the tire stiffness $C_K \leq 1442.64$ 
     (negative $g_4$).  
Function $g_3$ does not contribute to failure for any realistic combination of inputs, and there is a~possibility of $g_1$ turning negative when the damping force coefficient $K$ happens to have an extremely small positive value due to its appearance in the denominator. Note that the problem formulation in \cite{doi:10.1504/IJVD.1984.061093:suspension_3D,RASHKI201424:suspension_3D} is such that negative $K$ does not imply negative $g$. This possibility is manifested by the emergence of three failure events associated with low values of $K$, see the top right part of Fig.~\ref{fig:suspension_3D}.
If we disregard this negligible contribution of low positive values of $K$ to the failure probability and also the possibility of $g_3<0$, we are left with two parts of the failure surface formed by $g_2=0$ and $g_4=0$, i.e.,  two orthogonal planes with the following design points in the standard Gaussian space:
    $\beta_1 = \lvert 1442.64-1475.5503 \rvert /10 \approx 3.29$, 
    $\beta_2 = \lvert 391.212-431.7221 \rvert/10  \approx 4.05$. 
Since there is no interaction between these two failure modes, the simple FORM analysis of a~system failure probability yields
$p_{\mathrm{f,FORM}}
= \Phi(-\beta_1) + \Phi(-\beta_2) 
- \Phi(-\beta_1)   \Phi(-\beta_2) 
= 5.25001 \cdot 10^{-4} $. 






The proposed global sensitivities to the two dominating variables $C$ and $C_K$ are
 $ t_{C}^2 
    = \varphi(\beta_2)\Phi(\beta_1) /T 
    \approx 0.058$  
and 
$ t_{C_K}^2 
    = \varphi(\beta_1)\Phi(\beta_2) /T  
    \approx 0.942$, 
where $T = \varphi(\beta_1)\Phi(\beta_2) + \varphi(\beta_2)\Phi(\beta_1)$. 
Their respective values obtained from the numerical algorithm 
$
\pmb{t}^2
=
\{
t^2_C, t^2_{C_K}, t^2_K
\}
\approx
\{
0.057, 0.943, 8\cdot 10^{-6}
\}
$;
see~Fig.~\ref{fig:suspension_3D} top left. 

It is worth noting that the proposed algorithm can share all the calculated \gx\ outputs between different probability distributions $f_{\X}(\x)$ and may be suitable for a~reliability-based design optimization. For this example, however, we only took a~single particular case of input variable distributions listed in Tab.~\ref{tab:suspension_3D} from~\cite{RASHKI201424:suspension_3D}.

\subsection{Dynamic Response of a Non-linear Oscillator
\label{sec:oscillator}}

\begin{table}
\centering
\caption{Input variables for the non-linear oscillator example in Sec.~\ref{sec:oscillator}.}
\label{tab:oscillator}
\begin{tabular}{llcl}
  \toprule
Variable & Distribution  & $\mu$ & $\sigma$ \\
  \midrule
  $m$ & Normal  & 1 & 0.05 \\ 
  $k_1$ & Normal  & 1 & 0.1 \\
  $k_2$ &  Normal   & 0.1 & 0.01 \\ 
  $r$ &  Normal &  0.5 & 0.05 \\ 
  $t_1$ &  Normal & 1 & 0.2 \\ 
  $F_1$ & Normal  & 0.45 & 0.075 \\ 
\end{tabular} 
\end{table}

\begin{figure}[!htb]
    \centering
    \includegraphics[width=\textwidth]{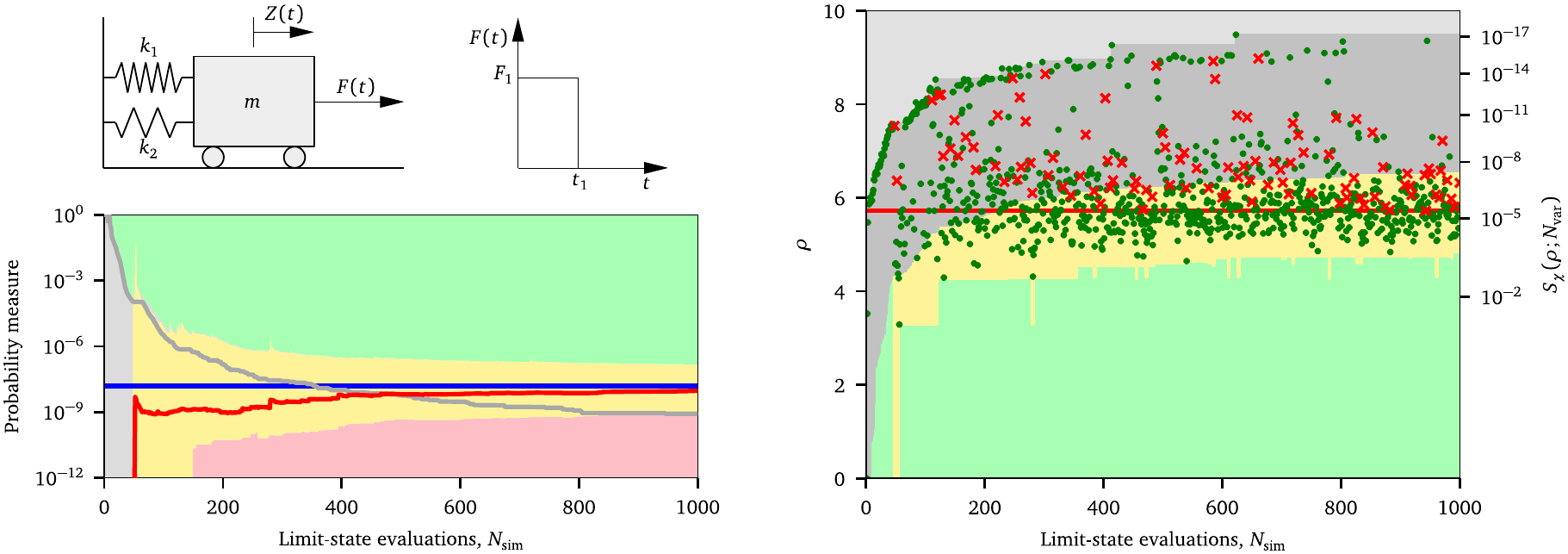}
    \caption{Performance of the method for the non-linear oscillator problem. Top left: illustration of a single degree spring mass oscillator. Left bottom: convergence diagram. Right: radial projection in standard Gaussian space. }
 \label{fig:nonlinear_oscillator}
\end{figure}

The following 6D LSF is inspired by the classic ``reliability benchmark'' in the field of structural reliability, which was studied in~\cite{BUCHER199057, Echard2013, WANG2021114172:fourbranch, Wang2022}. A non-linear undamped single degree of freedom system is illustrated in Fig.~\ref{fig:nonlinear_oscillator}. The limit state is defined by
\begin{equation}
     \label{eq:nonlinear_oscillator}
     g\left( \x \right) =
    3 r_y -  \left| \frac{2F_1}{m\omega_0^2} \sin \left( \frac{\omega_0t_1}{2}  \right)  \right|,
\end{equation}
where $\omega_0=\sqrt{(k_1+k_2)/m}=\sqrt{\Tilde{k}/m}$, $r_y$ is the displacement at which one of the springs yields, and the term in the absolute value parenthesis is the maximum displacement response. Here, we employed a~dimensional reduction by replacing the sum of two independent normal random variables representing the sum of spring stiffnesses with a single random variable $\Tilde{k} \sim N(\mu_{k_1}+\mu_{k_2},\, \sigma_{k_1}^{2}+\sigma_{k_2}^{2} ) \sim N(1.1,\, 0.0101) $. Therefore, we actually perform an analysis of a~5D problem.

The original formulation of the problem leads to a~very high failure probability ($\approx 3\cdot 10^{-2}$) and, therefore, cannot be considered a realistic benchmark of reliability methods. In the modification presented in this paper, the mean value and the standard deviation of the random force $F_1$ are modified to decrease the  failure probability to the order of $10^{-8}$. The parameters of independent Gaussian variables are presented in Tab.~\ref{tab:oscillator}. We employed the importance sampling method with  $10^7$ LSF evaluations to estimate the exact failure probability accurately: $\pF \approx 1.57 \cdot 10^{-8}$. 
The proposed algorithm yields 
the estimation of failure probability $ 0.941 \cdot 10^{-8}$ after the evaluation of 
$\Ns=1000$ ED points; see Fig.~\ref{fig:nonlinear_oscillator} bottom left.

Apart from the usual convergence of estimations presented in Fig.~\ref{fig:nonlinear_oscillator} bottom left, we present the radial distances $\rho$ of individual ED points in Fig.~\ref{fig:nonlinear_oscillator} right. The background along with the secondary vertical axis on the right-hand side provides an additional insight of what kind of event can be expected at the given radius $\rho$. The secondary vertical axis corresponds to the primary vertical axis: it shows the probability content inside an \Nv-ball with a~radius $\rho$ computed via Eq.~\eqref{eq:chi-cdf}.
The yellow band in the background shows  bounds of radial distances of ED points (vertices) of all the existing mixed simplices.  
The dark grey band, which may cover a~part of the yellow band, shows the gap between the radial distances $r$ and $R$, i.e., the radial range of a~Gaussian annulus between the most distant ED point and the nearest node outside the convex hull; see Fig.~\ref{fig:integration} right.
Fig.~\ref{fig:nonlinear_oscillator} right highlights the gap between the radial distance at which the ED points are placed, and the critical radial distance $r$, which corresponds to the offset of the nearest bounding hyperplane of the convex hull.
After the evaluation of  $\Ns=1000$ points, many of which are located at a~radial distance as large as 9, we still have the critical radius of only $r=6.58$. Not only the nearest facet is this close to the origin. In spaces of large dimensions, the convex hull constructed using many points on the surface of an \Nv-ball with a~large radius $R$ have much lower distances of the facets from the origin. The corresponding probability outside the convex hull may easily be several orders of magnitude greater than the probability outside the circumscribed \Nv-ball. If we use some other surrogate model, everything outside the convex hull could be treated as a~potentially erroneous extrapolation.





The global sensitivities obtained from the numerical algorithm are based on $\Ns = 1000$ ED points forming  
40481 mixed simplices. These simplices were cut by  1835 unique hyperplanes giving different normals. Numerical evaluation yields $
\pmb{t}^2
=
\{
t^2_m, t^2_{\Tilde{k}}, t^2_r, t^2_{t_1}, t^2_{F_1}
\}
\approx
\{
0.0069, 0.0836, 0.4526, 0.2464, 0.2105
\}
$. 
The largest group of 32009 mixed simplices is cut by a~single hyperplane with the following normal (in standard Gaussian space): $\pmb{n} \approx \{ -0.0732,  -0.2698,  -0.6661, 0.5158, 0.4605\} $. This normal roughly corresponds to the above-stated sensitivities. In total, the hyperplane separates 843 ED points which form vertices of the respective simplices. The offsets of planes which strictly separate failure and safe points are in the range of $(5.53995, 5.5413)$. A~rough estimation of the failure probability based on a~``middle'' plane can be obtained in the FORM fashion as  
$\Phi(-5.54) = 1.512 \cdot 10^{-8}$, which is quite close to the large sample analysis above. The geometrical insight provided by the proposed method reveals that the failure surface is almost linear (separation plane with a given position and known normal) and, therefore, the effective dimension of the 6-dimensional LSF after rotation of coordinates is one.

\section{Higher-dimensional Gaussian Space
    \label{sec:HD:testcases_nd}
}

The developed methodological framework has proved to be very robust and efficient in the case of low-dimensional problems. However, an immediate question arises as to how well the method scales with the problem dimension, because medium- and high-dimensional problems are of great interest for practical applications, too. 
A clear limitation stems from the computational geometry needed to maintain the structure of simplices. The number of geometrical objects associated with each simplex (nodes, lines, planes, hyperplanes, etc.) grows fast with the domain dimension; therefore, also the computational complexity grows exponentially. 
As will be shown in this section using numerical examples, the method is applicable to problems of the dimension $ \Nv \leq 8$. Limit state functions that have no chance of reducing the problem dimension to eight are not effectively analyzed by the proposed approach, which relies on exact geometry.

\begin{figure*}[!t]
    \centering
    \includegraphics[width=\textwidth]{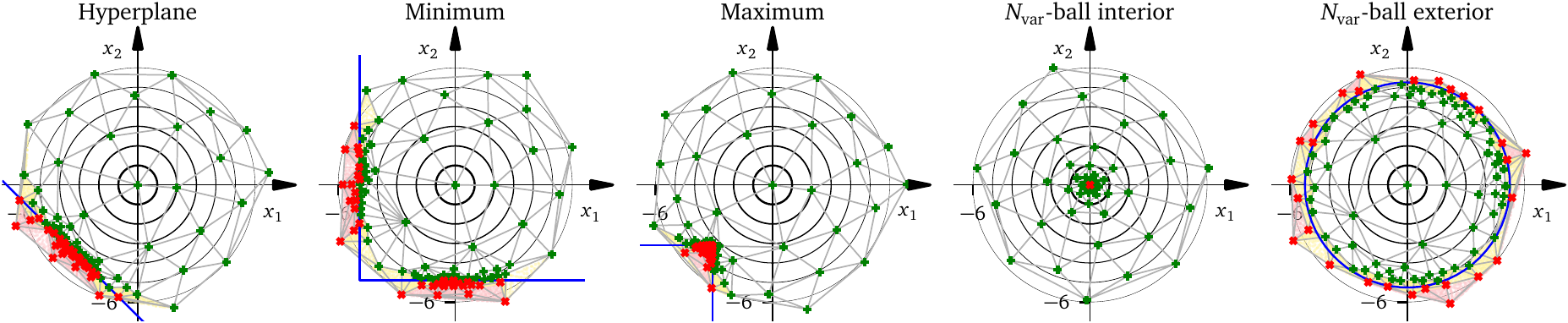}
    \caption{
           Illustration of the state after adaptive sequential addition of $\Ns=100$ for the five test cases in the case of $\Nv=2$ dimensions. 
    }
    \label{fig:testcases_nd_points}
\end{figure*}

In this section, we present five different limit state functions $\gx$ which are leading to geometrically or topologically similar failure surfaces in Gaussian spaces of arbitrary dimensions. The failure surfaces are composed of either (hyper)planes or (hyper)spheres, and they can be interpreted as functions representing a~system with additive independent components, or series (weakest-link) systems, systems with parallel components, or rotationally invariant problems. All the functions are defined such that the analytical solutions exist, and the failure probability is adjusted to exactly $\pF = 10^{-6}$. Fig.~\ref{fig:testcases_nd_points} illustrates the failure surfaces for the dimension $\Nv=2$. 

\subsection{Linear Failure Surface
    \label{sec:HD:testcases_nd:hyperplane}
}
The sum of random variables that forms a~single hyperplane failure boundary is the most common and the most often studied example in the field of structural reliability
\begin{align}
     g_{\mathrm{lin}}\left( \x  \right)
    &= \sum_{v=1}^{\Nv} x_v - C_{\mathrm{lin}} \, ,
\end{align}
where the constant 
$C_{\mathrm{lin}} = \Phi^{-1}\left( \pF \right) \cdot \sqrt{\Nv}$
is the product of the $\beta$ index with the square root of the domain dimension.
The failure probability has a~trivial solution as the probability content of a~part of the Gaussian density separated by a~linear boundary can be transformed into an univariate problem based on the most central failure $\beta$ point. Gradient-based methods have no difficulties locating the design point and forming the approximation of the failure surface.  
The values of the $g_{\mathrm{lin}}$ function provide reliable information about the distance to the failure surface.

The proposed method works with binary information, only and the refinement of the approximated failure surface is performed using ``mixed'' simplices. The failure surface is linear (a hyperplane), and also the boundaries of individual simplices are linear, i.e., formed by hyperplanes. Therefore, in this particular case, whenever there is a~simplex with vertices of the same classification, the whole simplex inherits the correct classification, because the failure is linear. The convergence plots for various domain dimensions are displayed in the leftmost column of Fig.~\ref{fig:testcases_nd}. It can be seen that the probability content in mixed simplices at fixed \Ns\ grows an increasing domain dimension, \Nv.

\subsection{Minimum Function -- a~Series System}

Consider a~system of \Nv\ independent and identically distributed components, which fails when \emph{any} of its \Nv\  components fails. The weakest component in such a~series system then becomes decisive and its performance must be compared to a~threshold
\begin{align}
    g_{\min}\left( \x  \right)
    = \min_{v=1,\ldots,\Nv} \left( x_v \right) - C_{\min}
    \, ,
\end{align}
where $ C_{\min}=  \Phi^{-1}\left( 1 - 
  \sqrt[\leftroot{-1}\uproot{3}{\Nv}]{1-\pF}
 \right)
$ is the ``safety threshold'' for each individual variable (we use high precision arithmetic evaluation of $C_{\mathrm{min}}$).

From all possible directions from the origin, only $1/2^{\Nv}$ are not pointing towards the failure region. 
The consequence of this arrangement is that the extent of the part of the failure surface, which has a~high probability density, is large and increases with the domain dimension. Therefore,  a~sufficient  refinement of the important part of the boundary needs high numbers of ED points in higher dimensions. Again, the probability associated with the mixed simplices decreases very slowly with an increasing number of LSF evaluations, see Fig.~\ref{fig:testcases_nd}.

\subsection{Maximum Function -- a~Parallel System}
Consider a~system of \Nv\ independent and identically distributed components, which fails when \emph{all} of its \Nv\ components fail. The failure of the system, therefore, occurs at the failure of the strongest component
\begin{align}
g_{\max} \left( \x  \right)
    = \max_{v=1,\ldots,\Nv} \left( x_v \right) -C_{\max}
    \, ,
\end{align}
where the threshold reads $C_{\max} =  \Phi^{-1}\left(\sqrt[\leftroot{-1}\uproot{3}{\Nv}]{\pF}\right)$.
The problem is, in a~way, a~counterpart to the previous $g_{\min}$ function.  The failure region is limited to just one of all the $2^{\Nv}$ ``quadrants'' (a~``black swan'' event). At the beginning, the extension algorithm must ``hit'' a~point from the failure region. This can necessitate many LSF calls as there is no clue about promising search directions and the space must be explored progressively from the origin towards all possible directions. To hit a~failure event for the first time is harder in high dimensions, because the number of potential directions is higher there, and it is a~matter of chance for which LSF evaluation  the first failure event occurs in the exploratory phase. This makes this problem different from the rest of the numerical examples, because there is a~greater scatter in the sample size for which the estimator provides the first answer. However, most of the failure probability is localized in the vicinity of a~single $\beta$ point, and once this failure ``corner'' is revealed, the refinement of the boundary via the triangulation near that point requires only a~small number of LSF evaluations. The estimators then quickly provide an accurate answer and the ``mixed'' simplices occupy a~very small probability content, see  Fig.~\ref{fig:testcases_nd}.


\subsection{\Nv-ball Interior}
Consider a~somewhat exotic situation in which the mean values (Gaussian space origin) lead to failure. The origin is the central point of a~single failure region in the shape of an \Nv-ball with radius $r_{\mathrm{int}}$
\begin{align}
    g_{\mathrm{int}}\left( \x \right)
    =  \sum_{v=1}^{\Nv} x_v^2  -  r_{\mathrm{int}}^2
    \, ,
\end{align}
where $r_{\mathrm{int}} = r(p_{\mathrm{in}}; \Nv) $ is set using Eq.~\eqref{eq:invFchi} such that the interior probability $p_{\mathrm{in}} = \pF =  10^{-6}$.
Deviations from the mean value which exceed $r_{\mathrm{int}}$ in any direction are safe.

With this definition of the LSF, the situation is somewhat untypical, because the failure region is not spreading to infinity, but it is bounded and very small in volume. Such a~localized failure region has the smallest possible extent of boundary, given the probability $\pF$. In this sense, it is a counterexample, and the proposed algorithm works only because the very first LSF evaluation is located in the failure region. Therefore there is no risk of encapsulating the failure region inside supposedly safe simplices. 

The standard approach of \IS\ or AS, in which the sampling density is inflating the variance compared to the original density, is even less efficient than the crude Monte Carlo. FORM and SORM methods are unable to provide reasonable \pf\ estimations at all. The proposed extension algorithm is successful in refining the triangulation around the point-like failure region. In the standard setting, the algorithm is quite aggressive, meaning that there are many exploratory points located quite far from the origin. Nevertheless, the estimator converges very rapidly towards the true failure probability; see Fig.~\ref{fig:testcases_nd}.


\subsection{\Nv-ball Exterior}
Consider a~situation in which failure occurs if the distance of a~point from the mean value exceeds a~predefined radius $r_{\mathrm{out}}$
\begin{align}
    g_{\mathrm{out}}\left( \x \right)
    = r_{\mathrm{out}}^2 - \sum_{v=1}^{\Nv} x_v^2
    \, ,
\end{align}
where the radius $r_{\mathrm{out}} = r(p_{\mathrm{out}}; \Nv)  $ is set using Eq.~\eqref{eq:invSchi} such that the exterior probability $p_{\mathrm{out}} = \pF=  10^{-6}$. 

\begin{figure*}[!htb]
    \centering
    \includegraphics[width=\textwidth]{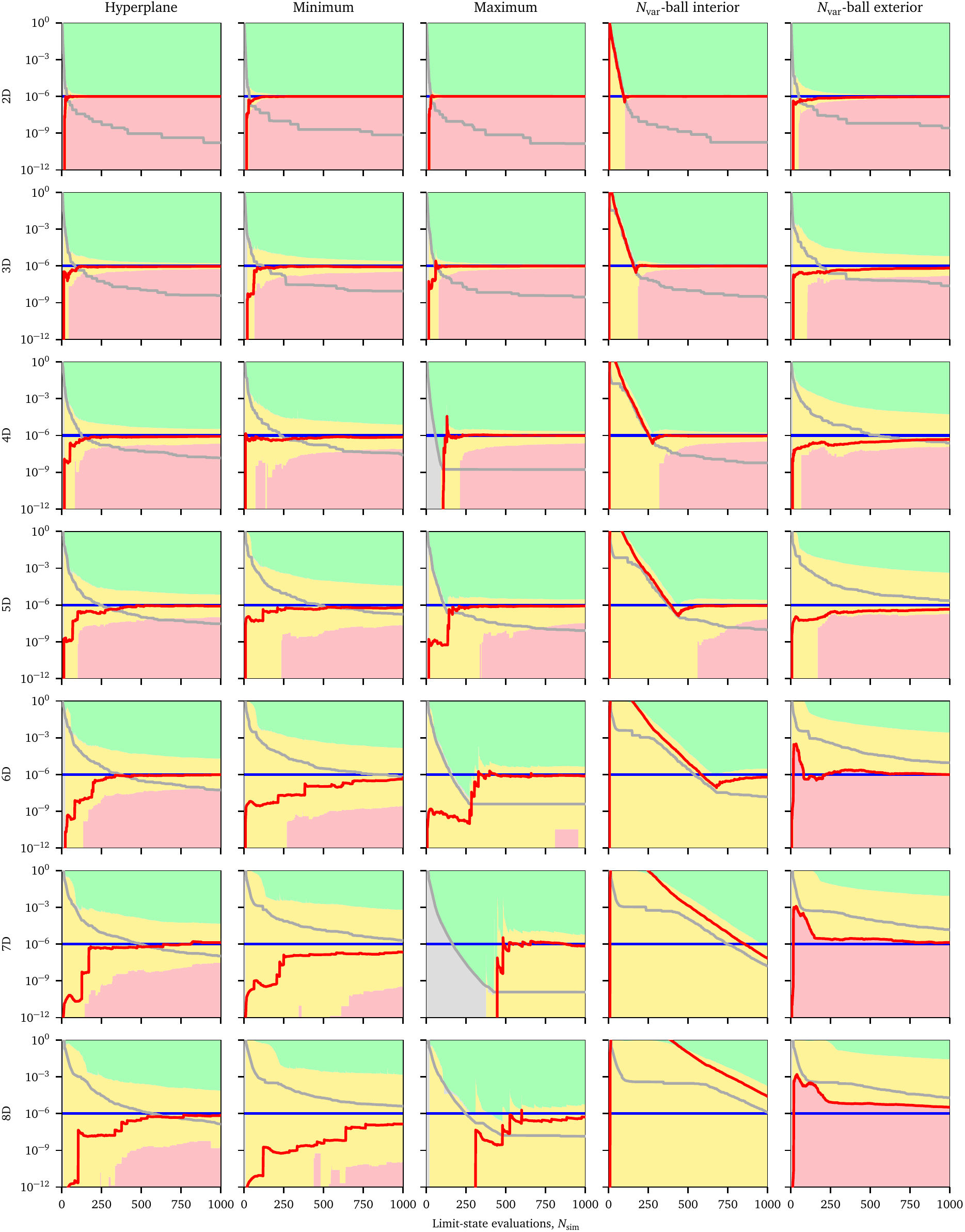}
    \caption{
            Convergence diagrams for general dimension problems. 
    }
    \label{fig:testcases_nd}
\end{figure*}

In the context of failure probability estimation and failure surface refinement, this is the most difficult scenario for the presented algorithm. The reason is that the extent of the failure surface is maximal for the given failure probability. All points on the failure surface are equally probable and the refinement must occur with the same density of points all over the surface of the \Nv-ball with radius  $r_{\mathrm{out}}  $. FORM and SORM struggle, because there is no unique ``design point''. When the design point search finds any point on the failure surface, FORM underestimates the true failure probability by setting $\pF = \Phi^{-1}( r_{\mathrm{out}})$, which is on the unsafe side. The right-hand side column in Fig.~\ref{fig:testcases_nd} demonstrates that the probability in ``mixed'' simplices decreases only very slowly in higher dimensions.

\subsection{Discussion of Results}

\begin{figure}[!htb]
    \centering
    \includegraphics[width=9cm]{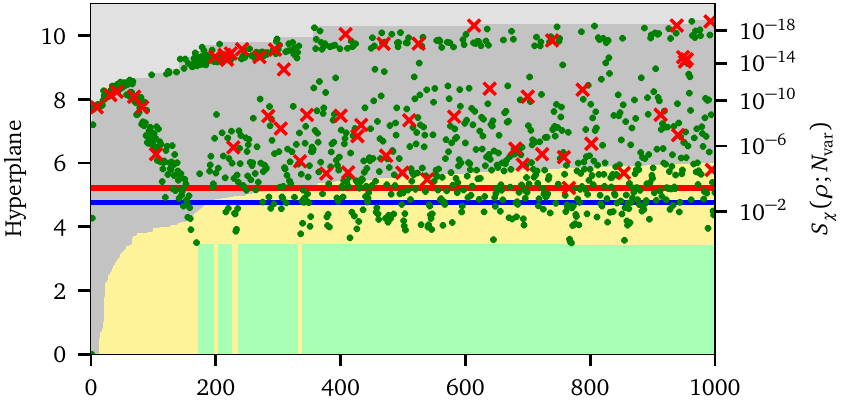}
    \includegraphics[width=9cm]{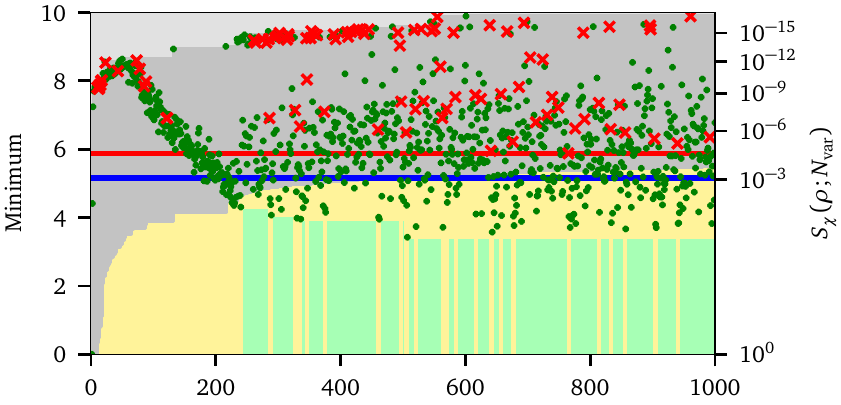}
    \includegraphics[width=9cm]{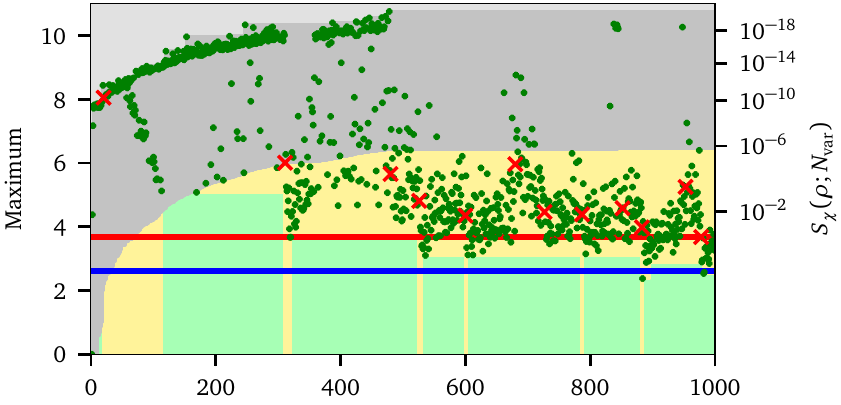}
    \includegraphics[width=9cm]{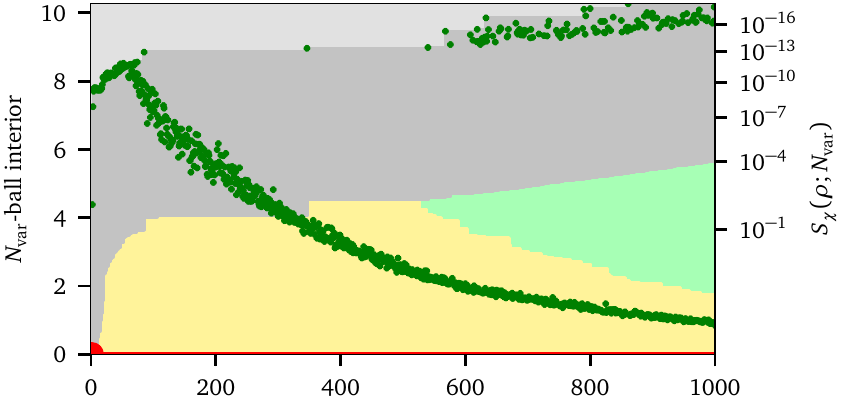}
    \includegraphics[width=9cm]{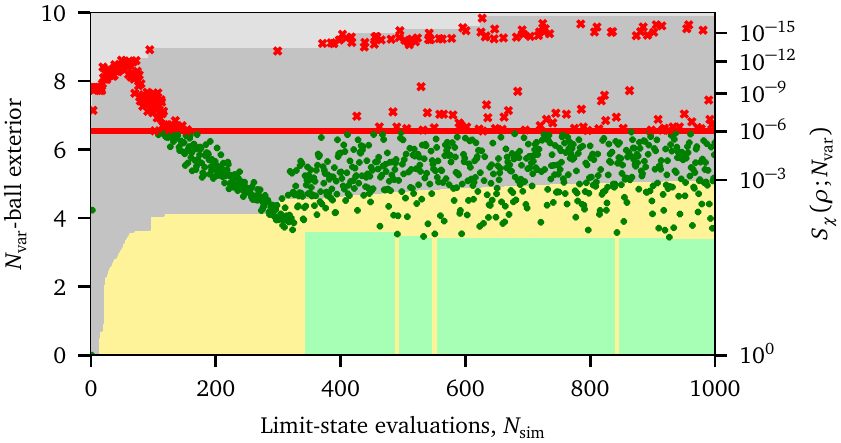}
    \caption{Radial graphs for the respective 8D problems
    }
 \label{fig:testcases_8d_graph}
\end{figure}

The behavior of the algorithm in eight-dimensional space is captured for all five functions in   
Fig.~\ref{fig:testcases_8d_graph}. The figure presents radial distances of individual ED points in the same way as previously reported in Fig.~\ref{fig:nonlinear_oscillator}.

As already discussed in the introductory part to this section, the exact geometry behind the construction of the ``scaffold'' made of simplices is intractable for a~dimension greater than eight on contemporary computers. 
The typical situation in higher-dimensional spaces with up to hundreds of ED points is that most of the convex hull is occupied by mixed simplices. It is so, because the purely safe or failure simplices need to have all $\Nv+1$ vertices of the same outcome type, and this is harder to achieve than in low-dimensional spaces. 

On the other hand, the proposed placement of new ED points into the centers of circumscribed circles/balls of the heaviest mixed simplices produces very good space-filling designs. When the LSF function is smooth, these ED points provide an excellent support for smooth surrogate models. While estimating failure probabilities with binary information becomes less efficient in higher dimensions, the use of smooth surrogate models with our ED points represents an excellent gain in efficiency.




\section{Dependent and/or Non-Gaussian Marginal Distributions
    \label{sec:spaces}
}

\begin{figure*}[!b]
    \centering
    \includegraphics[width=\textwidth]{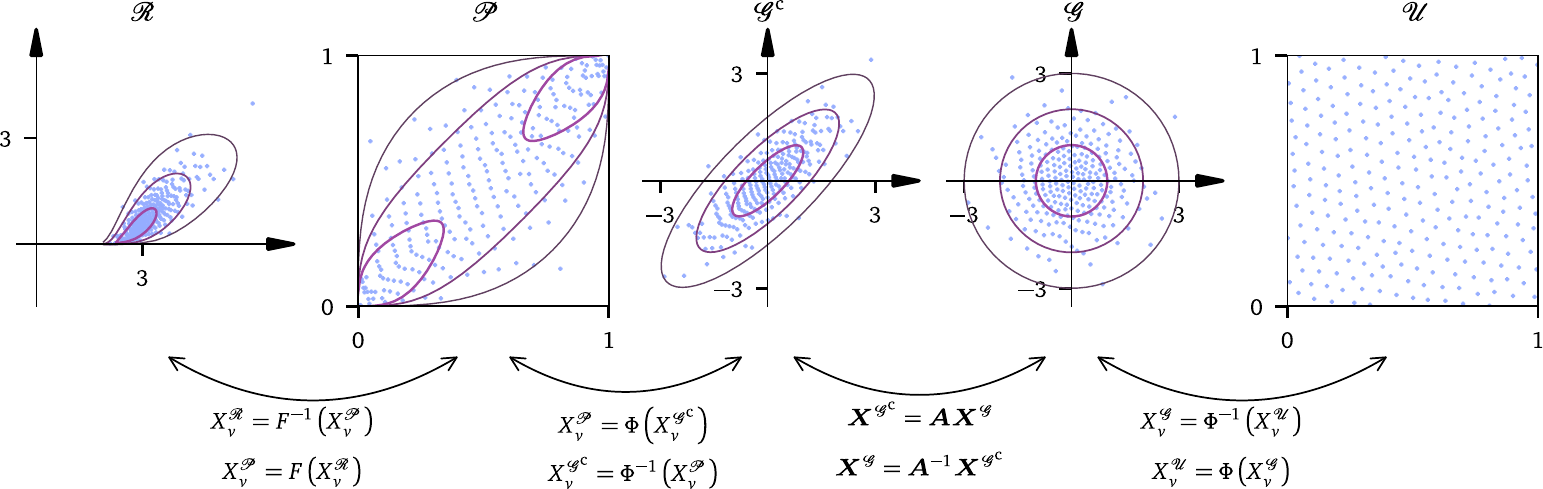}
    \caption{
           The chain of \emph{spaces} and \emph{transformations} involved in the Nataf model and considered in the paper.
    }
    \label{fig:spaces}
\end{figure*}

Until this point, the analysis was limited to independent normal input random variables. Similarly, most of the important results in the reliability theory and the related advanced tools, including classic FORM/SORM reliability methods, are available for these jointly Gaussian random vectors.
However, many real engineering applications feature correlated and/or non-Gaussian random variables.
There are two popular solutions for the construction of the joint probability density: the Rosenblatt transformation via the product of conditional distributions \cite{Rosenblatt1952,MelchBeck:2017} and the Nataf model \cite{Nataf,DerKiureghian1986} (sometimes referred to as the Gaussian copula approach \cite{Nelsen:Copulas:06}).
The purpose of this section is to demonstrate the use of Nataf transformation (see Fig.~\ref{fig:spaces}) with the proposed technique.
The joint distribution as well as the limit state function $\gx$ is defined in a~potentially non-Gaussian space with correlated marginals, which is called the real, original or \emph{physical space}, \RSp; see the illustration in the left part of Fig.~\ref{fig:spaces}. In this space, the state of the system is evaluated, because LSF is defined here. The probabilistic dimension of this space is equal to the number of input random variables, \Nv.

To the right of the \RSp\ space, we depict the space of copulas (dependence patterns among the cumulative distribution functions), i.e., space $\mathcal{P}$ with $\Nv$  uniformly distributed marginals over intervals $\left[0, 1 \right]$. The component-wise transformations are performed via the known distribution functions of individual marginals. The distribution functions in  $\mathcal{P}$ space provide a~simple means to map them individually to the space of dependent/correlated standard Gaussian variables \GcSp via a~translation process, sometimes called the \emph{memoryless iso-probabilistic transformation}. In this space, the Nataf model has the modified correlations between pairs of $\Nv$ underlying Gaussian variables, which must be solved numerically by solving simple bivariate integrals and for which some semi-empirical formulas are known \cite{DerKiureghian1986}. In case the correlation matrix in \GcSp\ obtained for all pairs is a valid positive definite matrix, the vectors from the correlated standard Gaussian space \GcSp\ can be mapped to the space of standard uncorrelated Gaussian marginals \GSp.

 Unlike all the previous transformations that were both component-wise and uniquely defined, decorrelation is non-separable and could be performed in an infinite number of ways -- even if we limit ourselves to linear maps; see \cite{NovVor:Trans:18}. We adopt the frequent assumption of linear mapping between \GcSp\ and \GSp, and from all possible such maps name two frequently used maps in the literature, namely the eigendecomposition (known as the Principal Component Analysis, proper orthogonal decomposition, Karhunen-Lo\`{e}ve expansion, orthogonal transformation of covariance matrix), and the Cholesky decomposition. Our current implementation uses Cholesky decomposition in which the first component, $x_1$, is identical in both Gaussian spaces, and all the other components are linear combinations of the preceding ones. We remark, though, that the principal component analysis may be a~preferred alternative as it provides an efficient way of reducing the dimension \GSp\ compared to the original dimension \GcSp\ by ignoring the components with the smallest variance contributions (eigenvalues)~\cite{NovVor:Trans:18,Vorech:08:CrossCorr}. This dimensionality reduction from $\Nv$ to $N_\mathrm{r}$ may be a~key to a~better effectiveness of the proposed method as well as for many other methods.

The last space presented is the \USp\ space of uniform independent marginals; see Fig.~\ref{fig:spaces} right. This is the space having an \emph{independence copula}. The coordinates are obtained by computing the inverse Gaussian distribution function of individual marginals in \GSp\ space. Any random vector is, therefore, following the uniform unit probability density over a~unit hypercube $\mathcal{U} \equiv \left[0,1 \right]^{N_\mathrm{r}}$. 

A straightforward solution for non-Gaussian correlated random vectors is to use the uncorrelated Gaussian space \GSp\ for all the tasks: the spatial decomposition, computation of cubatures, and the selection of the next ED point for an LSF evaluation. Sensitivities should be computed in the real space in which the individual variables have their clear meaning.
The following numerical example illustrates the option of using Nataf transformation to gain access to \GSp.

\begin{figure}[!htb]
    \centering
    \includegraphics[width=8cm]{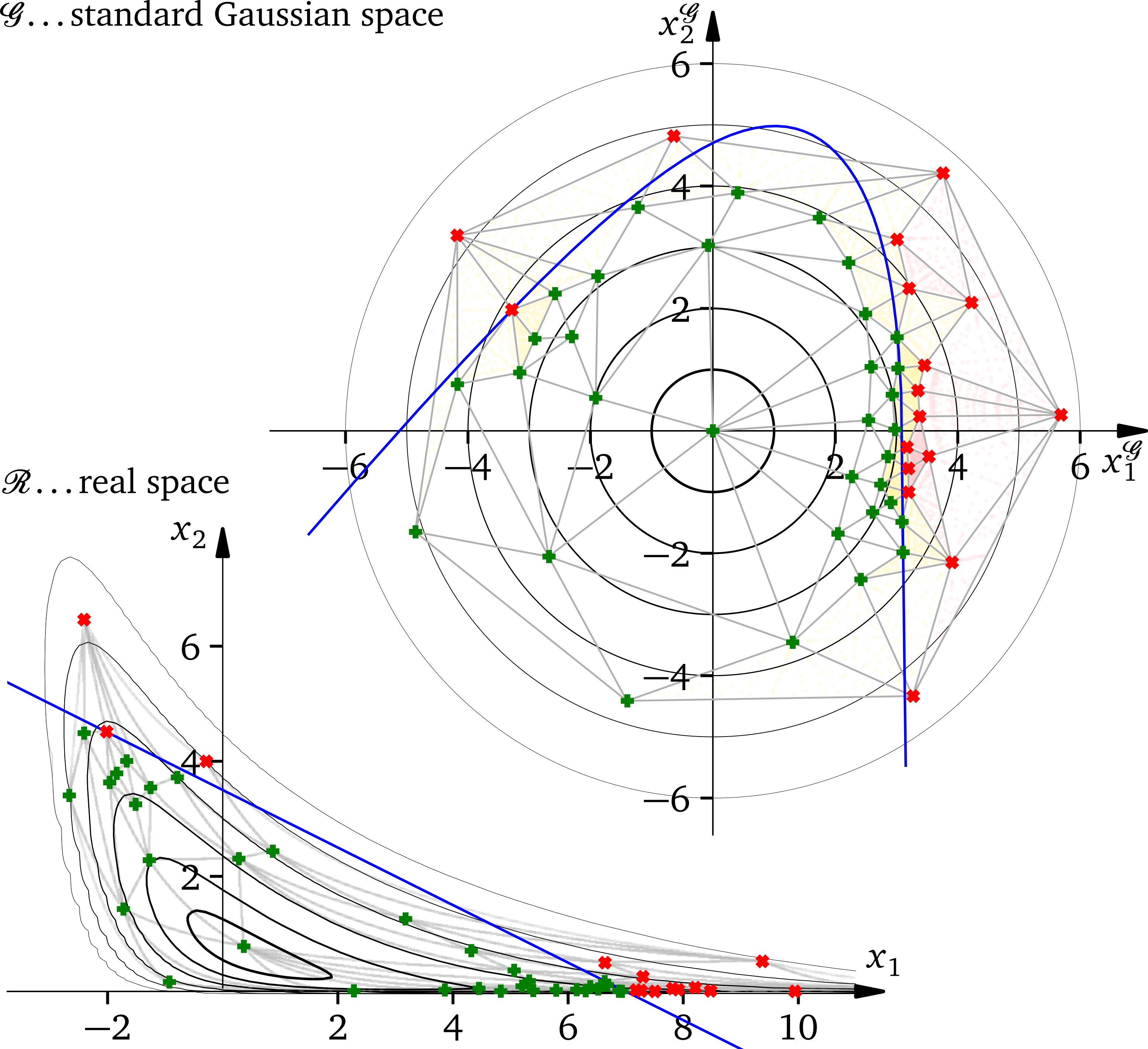}\\
    \includegraphics[width=9cm]{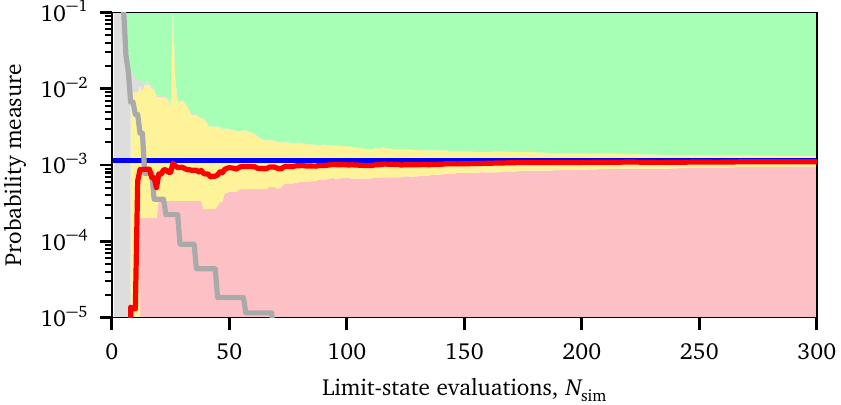}
    \caption{State classification in $\Ns=50$ ED points and the corresponding space decomposition 
             for the problem given in Eq.~\eqref{eq:nataf:example}.
        Top: standard Gaussian space \GSp.
        Middle: original (real) space \RSp.  
        Bottom: convergence diagram. 
        }
 \label{fig:nataf:natafm_plane_2D}
\end{figure}

Consider now a~linear LSF involving two correlated non-Gaussian variables
\begin{equation}
 \label{eq:nataf:example}
    g \left( x_1 ,x_2 \right)
    = 7 - x_1 - 2x_2.
\end{equation}
The marginal $x_1$ has Gumbel (right-skewed, i.e., ``max'') distribution known also as type I Fisher-Tippett distribution with the scale parameter equal to 1 and zero mode (location parameter). The second marginal $x_2$ has Weibull (``min'') distribution with unit scale parameter and shape parameter (exponent known as Weibull modulus) equal to $1.5$. This two-parameter Weibull variable is bounded from the left by the zero lower bound.
The Pearson correlation between these non-Gaussian variables equals $-0.708$ and failure events are signaled by $g<0$.

The true probability of failure cannot be uniquely determined, because the joint probability density $f_{\X}(\x)$ is not defined. However, we suppose that $f_{\X}(\x)$ constructed via the Nataf model is the true one.
The correlation  between the underlying jointly Gaussian vector obtained by the iso-probabilistic transformation of individual marginals equals $-0.8$. A~large-sample analysis in the Gaussian space \GSp\ provides the reference failure probability $\pF \approx 1.14 \cdot 10^{-3}$.
The comparison of the top and middle parts of Fig.~\ref{fig:nataf:natafm_plane_2D} provides an illustration of how the spatial decomposition in the standard Gaussian space \GSp\ (top) transforms to the original space \RSp\ (middle), and how strongly the linear failure surface in \RSp\ may get distorted in \GSp. The bottom part of Fig.~\ref{fig:nataf:natafm_plane_2D} documents the convergence of the estimation of failure probability towards the exact solution and the decay of outside and mixed probabilities.

The linear failure surface in the \RSp\ space corresponds to simple sensitivities $t^2_1 = 0.2$ and $t^2_2 = 0.8$. These values obtained by the algorithm in the \RSp\ space correspond to the slope of the line $1/2$, which has a~normal $\{1,2\}$. These sensitivities can be used to reduce the dimension of the original input variables.

The sensitivities in \GSp\ are dependent on the selected linear transformation between \GcSp\ and \GSp\ 
\cite{NovVor:Trans:18}. The Cholesky transformation used in this example has the $x_1$ identical in both spaces, while the principal component analysis would rotate the coordinates; compare with the same numerical example in \cite{VORECHOVSKY2022115606}. In our case, the rotation visible in Fig.~\ref{fig:nataf:natafm_plane_2D} top is such that the sensitivities to coordinates in this Gaussian feature space are quite unbalanced:
$t^2_1 = 0.993$ and $t^2_2 = 0.007$.
We remark that the transformation between \GcSp\ and \GSp\ may provide the possibility of dimensional reduction, which, for the case of linear transformation, is well defined via the eigenvalues in the principal component analysis. The proposed sensitivities in the feature space \GSp\ provide additional information about a~potential dimensional reduction of that space.


We remark that the rightmost space depicted in Fig.~\ref{fig:spaces} may seem to be an alternative space for the tasks described in this paper. One can be tempted to construct the spatial decomposition in \USp\ space with the argument that the volumes are directly equal to the probabilities associated with the subdomains and, therefore, the estimation part becomes easier. We would like to emphasize that the \USp\ space is particularly unsuitable for practical applications in which the failure events are associated with small probabilities, i.e., tiny territories in \USp\ which are located very close to the boundary of the unit hypercube \USp. The numerical description of these boundary territories is problematic.

\section{Conclusions}

The paper provides methods and algorithms for the estimation of rare event (failure) probability for computer models expressing the fitness of a~product depending on random input variables. This ``black box'' limit state function can be highly nonlinear, implicit, noisy, have discontinuities or providing simply a~binary output only, and its evaluation is expected to be very expensive. 
 Due to the use of computational geometry, the proposed approach is limited to small dimensions (not exceeding eight).

The methods of local decomposition of the design domain are not being used in the field of structural reliability, and the paper proposes to segment the  design domain into simplices.
The continuously expanded and refined geometrical structure enables an effective classification of the expected event inside it (a~surrogate model), and supports the decision about the location of the next limit state function evaluation and also the effective integration of the probability density with the ultimate goal of estimating  the failure probability.
The proposed adaptive algorithm is completely gradient-free as it only uses the categorical output from the performance function: failure, or success. This makes the algorithm very robust against noise and fluctuations in the analyzed function, and it is invariant under problem reformulations and reparameterizations. The algorithm automatically discovers multiple design points and discretizes separate failure surfaces proportionally to the probability density. The main difference from many existing statistical sampling-based methods is in the achieved 
approximate knowledge of the geometrical representation of the problem and its refinement. This information provides a~valuable insight into the structure of the problem, which has a great potential for a~better utilization of the precious pointwise information obtained from previous evaluations of the expensive computer model. Purely statistical strategies embodied in the simulation methods disregard the structure of the problem in the space of random variables.

The proposed algorithm focuses on two  tasks: (i) the \emph{extension} of the experimental design by adding new points to it such that the balance between the expansion of the geometrical structure and its refinement around the failure surface is maintained, and (ii) the \emph{estimation} of probabilities both inside the structure discretized into simplices and also in its exterior. The extension is performed sequentially by adding one point at a~time while adapting the structural decomposition to the newly obtained information.
The \emph{estimation} of probabilities is achieved by a~combination of deterministic cubature rules inside the simplices, analytical expression for a~region outside the hyperball containing the convex hull, and effective integration via the application of the divergence theorem.

Additionally,  new global sensitivities are defined, which reflect the state transition between safe and failure states along the failure surface. This makes them well-suited for binary limit state functions. They are computed as by-products of the geometrical structure constructed in the proposed framework.

\section*{Acknowledgement}
The work has been supported by an internal project no. FAST-K-21-6943 sponsored by the Czech Ministry of Education, Youth and Sports, and also by the ``Quality Internal Grants of BUT (KInG BUT)'' project, Reg. No. CZ.02.2.69/0.0/0.0/19\_073/0016948, which is financed from the OP RDE.

\section*{CRediT authorship contribution statement}

\textbf{Aleksei Gerasimov:}
Conceptualization,
Methodology,
Software,
Investigation,
Data Curation,
Writing -- Original Draft,
Visualization,
\textbf{Miroslav Vo\v{r}echovsk{\'{y}}:} 
Conceptualization,
Methodology,
Validation,
Writing -- Original Draft,
Writing -- Review \& Editing,
Supervision,
Visualization,
Project administration

\appendix
\addcontentsline{toc}{section}{Appendices}

\section{Determination of Convex Hull Geometry
\label{appendix:outside:convex_hull}
}

Convex hull (sometimes called convex envelope or convex closure) of points is the smallest convex set that contains it. 
Convex hull in an $\Nv$-dimensional Euclidean space is formed by vertices and hyperplanar facets (\Nv-1  dimensional simplices). 
Convex hull can, therefore, be seen as the intersection of half-spaces and can be constructed using a~set of 
hyperplanes only. In this paper, the \lstin{QHull} library (as part of the \lstin{SciPy} package) was used for the convex hull construction. 
The Quickhull algorithm  for spatial dimension $\Nv \geq 4 $ has the worst-case time complexity $O(n_\mathrm{p} n_\mathrm{f}/n_r)$, where $n_\mathrm{p}$~is the number of input points and $n_r$~is the number of processed points \cite{Barber96thequickhull}. Being directly proportional to the number of facets $n_\mathrm{f}$, Quickhull's complexity is  asymptotically $O(n_\mathrm{p}^{\Nv/2})$~\cite{SEIDEL1995115} (unlike in two and three dimensions, in which  Quickhull's worst-case time complexity is $O(n_\mathrm{p} \log n_r)$).
For high-dimensional problems, where the exact computational geometry analysis by \lstin{QHull} becomes intractable, convex hull approximations can be used~\cite{GerVor:ESREL2021}. 

Given that $\x_1, \x_2, \ldots, \x_{k}$ are the vertex coordinates of a~facet, the equation of the hyperplane in which the facet lies can be written as the determinant of the following matrix with the coordinates extended with the first row and a column of ones 
\begin{equation} 
     H = 
      \begin{vmatrix}
     \x, & 1\\ 
     \x_1, & 1\\ 
     \x_2, & 1\\ 
     \vdots & \vdots \\ 
     \x_{k}, & 1\\ 
 \end{vmatrix} = 0.
\end{equation}
This equation can be expressed in vector form as 
\begin{equation}
H = \bm{ax}+b =a_1x_1+a_2x_2 + \dots + a_k x_k + b = 0,
\end{equation}
where $\pmb{a}$ is the unit normal vector and $b$ is the offset from the origin. 
Every facet delimits the \textit{inside} and \textit{outside} of the convex hull. The ``inside'' part is limited by points which lie on the same side of all the facets. Suppose there are $n_f$ facets forming the boundary of the convex hull. Each facet has its unit outward normal, i.e., a~vector of $\Nv$ coordinates. These normals can be arranged in matrix  $\pmb{A}$ and the respective offsets can be stored in vector $\pmb{b}$ for all $n_f$ facets. A~point $\x^*$ lies inside the convex hull only if all the following inequalities hold
\begin{equation}\label{eqAxb}
    \pmb{Ax}^*+\pmb{b} \leq \pmb{0}.
\end{equation}
The radius $r$ of the ball \emph{inscribed} into the convex hull is simply $r = \min_{i=1,\ldots,n_f}( -b_i)$.

\section{Spherical Coordinates and $\Nv$-ball in Standard Gaussian Space}
\label{appendix:nvball}

In this section, useful tools based on the standard spherical coordinate  representation of the Gaussian space are reviewed.

The joint density at any point $\x = \{ x_1,\ldots,x_{\Nv} \}$ in the Gaussian space, which is the product of univariate Gaussian densities of the individual marginals, can be greatly simplified
\begin{equation}
    \label{eq:stdGauss:point:x}
    f_{\X}(\x)
    =
    \prod_{v=1}^{\Nv}  \varphi(x_i)
     \equiv \varphi(\rho;\Nv)
    =
    \frac{1}{(2\pi)^{\Nv/2}}\exp{ \left(-\frac{\rho^2}{2} \right) }
    ,
\end{equation}
where $\varphi(x) = {(2\pi)^{-1/2}}\exp{ \left(-{x^2}/{2} \right) } $ is the standard univariate Gaussian density, 
and $\rho$ is the univariate Euclidean distance of $\x$ from the origin:
 $   \rho =  \lVert \x \rVert  =  \sqrt{ \x \tran \x } = \sqrt{\sum_{v=1}^{\Nv} x_v^2}$ , 
    $v=1,\ldots,\Nv$.
    

An important geometrical entity in the proposed method is an \Nv-ball with radius $\rho$, which is centered at the origin of the coordinate system. The volume and the surface of the $\Nv$-ball, denoted as $B_{\rho}  \in \mathbb{R}^{\Nv}$, read
\begin{align}
\label{eq:vol}
    \mathrm{Vol} \left[ B_{\rho} \right]
    &=
    \frac{\pi^{\Nv/2}}{\Gamma\left(  \frac{\Nv}{2}+1 \right)}
    \rho^{\Nv},
    \\
\label{eq:sur}
    \mathrm{Sur}\left[  B_{\rho} \right]
    &=
    \frac{2 \pi^{\Nv/2}}{\Gamma\left(  \frac{\Nv}{2}\right)}
    \rho^{\Nv-1}.
\end{align}

Another important ingredient in the proposed algorithm is the distribution function for a~random distance $\rho$ in the Gaussian space. Assume a~randomly selected point $\x$. Its Euclidean distance from the origin has $\chi$ (chi) distribution with $\Nv$ degrees of freedom.
The probability density function of a~distance $\rho > 0$ reads
\begin{align}
\label{eq:chi-pdf}
    f_{\chi} (\rho; \Nv)
    &=   \frac{2^{1-\Nv/2} }
           { \Gamma\left(\frac{\Nv}{2} \right)  }
        \rho^{\Nv-1}
        \exp \left(-\frac{ \rho^2}{2} \right)  \\
    &= \varphi(\rho;\Nv)  \cdot \mathrm{Sur}\left[  B_{\rho} \right]
        ,
\end{align}
where $\Gamma(\cdot)$ is the standard (complete) gamma function.
The corresponding cumulative distribution function,
defined as
$ F_{\chi} (\rho; \Nv)  = \int_{0}^{\rho}  f_{\chi} (t, \Nv) \dd t$,
reads
\begin{align}
\label{eq:chi-cdf}
    F_{\chi} (\rho; \Nv)
    =
    \frac{2^{1-\Nv/2} }
    { \Gamma\left(\frac{\Nv}{2} \right)  }
    \int_{0}^{\rho}
    t^{\Nv-1}
    \exp  \left(-\frac{ t^2}{2} \right) 
    \dd t
    = P_{\circ} \left( \frac{\Nv}{2} , \frac{\rho^2}{2}  \right),
\end{align}
where $P_{\circ} ( s, x) = \gamma(s,x) / \Gamma(s)$ is the \emph{regularized lower incomplete gamma function}.
$P_{\circ} ( s, x)$ is one of the ``special functions'' and it is a~standard part of various mathematical libraries, such as \lstin{SciPy} \cite{2020SciPy-NMeth}, which is available in \lstin{Python} (function \lstin{gammainc($\cdot,\cdot$)}).

The probability content corresponding to the \emph{interior} of an \Nv-ball with radius $r$ reads
$p_{\mathrm{in}} (r,\Nv) = F_{\chi} (r; \Nv) \equiv P_{\circ}(\Nv/2,r^2/2)$.

The inverse of the regularized lower incomplete gamma function, $P_{\circ}^{-1}$,  can be used to compute the  \emph{radius} $\rho$ of a~ball $B_{\rho} $ that contains a~point with the prescribed probability $p_{\mathrm{in}}$
\begin{equation}
    \label{eq:invFchi}
    \rho (p_{\mathrm{in}}; \Nv) 
    \equiv
    F_{\chi}^{-1}(p_{\mathrm{in}}; \Nv) 
    =
    \sqrt{ 2 \, \left[ P_{\circ}^{-1}(\Nv/2, p_{\mathrm{in}} )\right] },
\end{equation}

where $F_{\chi}^{-1}(p_{\mathrm{in}}; \Nv) $ is the inverse distribution function of the $\chi$-distributed random distance.

We now treat the \emph{exterior} of \Nv-ball $B_{\rho} \in \mathbb{R}^{\Nv}$ with radius $\rho$ separately.
The probability of a~point falling outside the \Nv-ball with radius $\rho$ is
 \begin{align}
    \label{eq:prob:out}
    p_{\mathrm{out}} (r;\Nv) 
    &= 1 - p_{\mathrm{in}} (\rho;\Nv)  = 1 - F_{\chi} (\rho; \Nv) \\ \nonumber
    &= S_{\chi} (\rho; \Nv) \equiv Q_{\circ}(\Nv/2,\rho^2/2),
\end{align}
where $S_{\chi} (\rho; \Nv) = 1 - F_{\chi} (\rho; \Nv)$ is the \emph{survival function} of $\chi$ distribution with $\Nv$ degrees of freedom.
Even though one can compute the probability of the exterior of an $\Nv$-ball by complementing its interior to unity, this solution may be numerically unstable due to floating point arithmetic with a~finite precision. We recommend using another standard ``special function'': the \emph{regularized upper incomplete gamma function},  denoted here as $Q_{\circ}( \cdot, \cdot)$, see \lstin{gammaincc($\cdot,\cdot$)} in \lstin{SciPy}.


Analogously to Eq.~\eqref{eq:invFchi}, the inverse of the regularized upper incomplete gamma function, $Q_{\circ}^{-1}$,  can be used to compute the radius of $B_{\rho}$ with the exterior of the prescribed probability $p_{\mathrm{out}}$
\begin{equation}
    \label{eq:invSchi}
    \rho(p_{\mathrm{out}}; \Nv) 
    \equiv
    S_{\chi}^{-1}(p_{\mathrm{out}}; \Nv) 
    =
    \sqrt{ 2 \, \left[ Q_{\circ}^{-1}(\Nv/2, p_{\mathrm{out}} )\right] },
\end{equation}
where $S_{\chi}^{-1}(p_{\mathrm{out}}; \Nv) $ is the inverse survival function of the $\chi$ distribution.

\section{Divergence Theorem (Gauss-Ostrogradsky) for Probability Integration
\label{sec:divergenceTheorem}
}

The joint probability density function
$f_{\X}(\x) \equiv \prod_{v=1}^{\Nv} f_v(x_v)$
of vector $\X$ is jointly Gaussian with identical independent univariate marginal densities $f_v(x_v)$, $v=1,\ldots,\Nv$. Each variable $v$ has the corresponding cumulative distribution function $F_v(x_v)$.

\begin{figure*}[!tb]
    \centering
    \includegraphics[width=\linewidth]{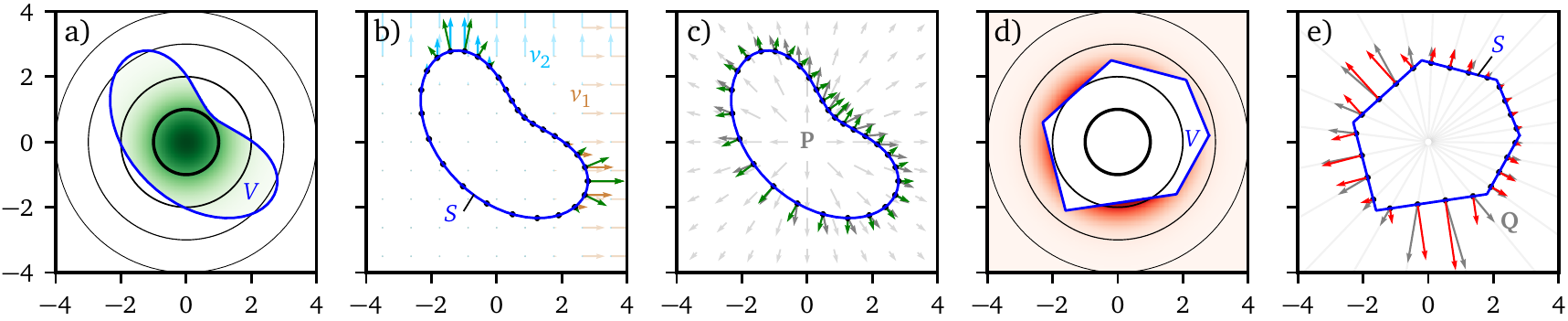}
    \caption{
            Applications of the divergence theorem to probability estimation.
            a) Probability inside a~closed region (spatial volume $V$).
            b) Two orthogonal vector fields and their dot product with normals of the surface $S$ enclosing the integrated volume.
            c) Radial vector field for integration in the interior probability.
            d) Probability outside a~closed region (spatial volume $V$).
            e) Radial vector field for integration in the outside probability.
    }
    \label{fig:divergence}
\end{figure*}

We propose using the divergence theorem to compute the probability inside a~closed region in $\DD$.
Given a~density $f(\x)$ and a~differentiable vector field $\vb{F}(\x)$ defined in \DD, the volume integral in subdomain $V$ can be transformed into a contour integral taken over the surface $S$ enclosing the volume $V$
\begin{align}
\label{eq:divtheorem}
        \iiint\limits_V    \div \left(  f(\x) \vb{F}(\x)  \right) \dd V 
        = 
        \oiint\limits_{S}  f(\x) \vb{F}(\x) \cdot \vb{n}(\x) \, \dd S
    \, ,
\end{align}
where $\vb{n}(\x)$ is the outward normal of the closed surface, see Fig.~\ref{fig:divergence}. The nabla operator dotted with the product of density and the vector field is a~scalar quantity, which is integrated on the left-hand side. This integral over the volume $V$ can be replaced by
an integral on the right-hand side, which is taken over the surface of dimension $\Nv-1$ only. The integrand is
the vector field whose components are projected onto (dotted with) the outward normal $\vb{n}$ and scaled by the local density $f(\x)$. The integral of the density in volume $V$
is effectively transformed to an integral over the surface which has a~lower dimension and, moreover, a wise selection of the vector field $\vb{F}(\x)$ can dramatically improve the accuracy of the numerical estimation via the right-hand side.

For cubature of probability inside a closed region $V$ located at a general position in Gaussian space, we propose a~vector field $\vb{F}_{\Nv}(\x)$ with $\Nv$ components constructed as ratios between the distribution function and density along each direction: 
$\vb{F}_{\Nv}(\x) = \left[ 
              \frac{F_1(x_1)  }{f_1(x_1)  },  
              \frac{F_2(x_2)  }{f_2(x_2)  }, \ldots, 
              \frac{F_\Nv(x_\Nv)}{f_\Nv(x_\Nv)}
       \right]$.
The surface integral in Eq.~\eqref{eq:divtheorem} then features \Nv\ components, which all have a simple form. Any $v$th component is a product of all densities, but the $v$th  one with the distribution function of $v$th variable:
    $F_v (x_v) \prod_{j=1, j\neq v }^{\Nv} f_v(x_v)$.
These components of the vector field are mutually orthogonal; see the light golden and the light blue vector fields $v_1(\x)$ and $v_2(\x)$ in Fig.~\ref{fig:divergence} b.

    The differentiation of such $v$th term with respect to $x_v$ yields $f_{\X}(\x)$.
    Therefore, the left-hand side (lhs) of Eq.~\eqref{eq:divtheorem} is \Nv\ times the probability content inside the volume $V$ and the corresponding right-hand side (rhs) splits into a sum of $\Nv$ simple contour integrals 
\begin{align}
\iiint\limits_V  
    \underbrace{\div \left(  f(\x) \vb{F}_{\Nv}(\x)  \right)}_
    {\Nv f_{\X}(\x) } \dd V 
   &= \Nv p_V 
   &\text{(lhs)}
    \\
    \oiint\limits_{S}  f(\x) 
    \underbrace{\vb{F}_{\Nv}(\x) \cdot \vb{n}(\x)}_{\sum_{v=1}^{\Nv} \left( \frac{F_v(x_v)  }{f_v(x_v)} n_v (\x)\right)}
    \, \dd S &=
    \oiint\limits_{S}   \sum_{v=1}^{\Nv} 
                    \left(
                    F_v (x_v) n_v(\x) \prod_{j=1, j\neq v }^{\Nv} f_j(x_j)
                    \right)
                    \, \dd S 
                    = \Nv p_V 
                       &\text{(rhs)}
\end{align}
This computation considers all \Nv\ orthogonal components, and the final probability $p_V$ is obtained by taking the result of the rhs and dividing it by \Nv. One can also simplify the vector field by considering only a subset of nonzero components. The simplest case is a single component for the selected direction $v$:
$ 
\left[ 
              0,  \ldots, 
              \frac{F_v(x_v)  }{f_v(x_v)  }, \ldots, 
              0
       \right]$.
The divergence of such a~vector field yields a~nonzero contribution from the $v$th direction only and, therefore, the left-hand side of Eq.~\eqref{eq:divtheorem} simplifies to 
   $ \iiint\limits_V     f_{\X}(\x) \dd V  =  p_V$.   
   The corresponding right-hand side becomes
\begin{align}
   p_V
  =
    \oiint\limits_{S}  f(\x)  \left( \frac{F_v(x_v)  }{f_v(x_v)} n_v (\x)\right)\, \dd S 
    .
\end{align}

When the centroid of a closed region $V$ is located approximately in the origin and the boundary is approximately spherical, as visible in Figs.~\ref{fig:divergence} c and d, the probability content inside $V$ can be better integrated by another differentiable vector field $\vb{v}$. A slightly reformulated version of the divergence theorem reads
\begin{align}
\label{eq:divtheorem2}
        \iiint\limits_V    \div \vb{v}(\x) \dd V 
        = 
        \oiint\limits_{S}  \vb{v}(\x) \cdot \vb{n}(\x) \, \dd S
    .
\end{align}
A suitable selection for the vector field $\vb{v}(\x)$ is a rotationally symmetrical divergent field $\vb{P}(\x)$, in which the direction is the unit radius vector $\vb{P}(\x)$ at point $\x$, and the intensity is also given by the distance from origin, $\rho$
\begin{align}
\vb{P}(\x) = \frac{ F_{\chi} (\rho; \Nv)} 
               {\mathrm{Sur}\left[  B_{\rho} \right]} 
               \vec{\rho}(\x),
\end{align}
where $F_{\chi} (\rho; \Nv)$ is the distribution function defined in Eq.~\eqref{eq:chi-cdf}.
Vector field $\vb{P}(\x)$ is visualized in 
   Fig.~\ref{fig:divergence}~c using light gray arrows.
At any point, the vector field has only a~ radial component and its intensity reads 
${ F_{\chi} (\rho; \Nv)}/{\mathrm{Sur}\left[  B_{\rho} \right]} $.
Using the spherical coordinate system in Eq.~\eqref{eq:divtheorem2}, one can show that both sides are equal to the probability content $p_V$ inside the volume $V$.
We now show that this choice of the vector field can be directly utilized for the cubature. An important step in the analytical derivation is to note that the divergence of a radial vector field, $\div \vb{v}(\x)  $, in the spherical coordinate system, is equal to
\begin{align}
\label{eq:div_spherical_coord}
         \div \vb{v}(\x)  
        & = 
        \frac { 1 }{ \rho^{\Nv-1} }
        \frac{\partial }{\partial \rho}
        \left( \rho^{\Nv-1}  \vb{v}_\rho(\rho) \right) ,
\end{align}
where $\vb{v}_\rho(\rho)$ is the radial component of $\vb{v}(\x)$ independent of the angular coordinate. In our case of $\vb{v}(\x) \equiv \vb{P}(\x)$, it is the only component, and the substitution of $\vb{P}(\x)$
into Eq.~\eqref{eq:div_spherical_coord} reveals that Eq.~\eqref{eq:divtheorem2} represents the probability content inside $V$.

In the method proposed in this paper, it is needed to accurately estimate the probability \emph{outside} a closed region $V$. In particular, the closed region is a convex hull bounded by linear boundaries (line segments, planes, hyperplanes) and the outside probability is very small; see the illustration in Fig.~\ref{fig:divergence}~d. The convex hull is bounding the set of \Ns\ ED points at a given stage of sampling, and the probability content outside it is denoted $\PO{\Ns}$.
In such a case, it is better to use yet another vector field in Eq.~\eqref{eq:divtheorem2}, which is, in a way, complementing  the rotationally symmetrical field $\vb{P}(\x)$. The only difference is that the intensity is computed from the complement to $F_{\chi} (\rho; \Nv)$, that is, $S_{\chi} (\rho; \Nv) = 1 - F_{\chi} (\rho; \Nv)$. The vector field reads
\begin{align}
{\vb{Q}}(\x) = \frac{ S_{\chi} (\rho; \Nv)} 
               {\mathrm{Sur}\left[  B_{\rho} \right]} 
               \vec{\rho}(\x)
               .
\end{align}
Plugging this vector field, in which $\rho$ is the distance of any point $\x$ from the origin, into Eq.~\ref{eq:divtheorem2} yields the expression for the evaluation illustrated in Fig.~\ref{fig:divergence}~e
\begin{align}
\label{eq:divtheorem:P:outside}
        \PO{\Ns} = 
        \oiint\limits_{S}  \vb{Q}(\x) \cdot \vb{n}(\x) \, \dd S.
\end{align}
In practical computation, the individual faces forming the boundary $S$ of the convex hull are integrated separately by using the very same deterministic cubature rules as described in Sec.~\ref{sec:estimation:inside}. The only difference from the use of cubatures for \Nv-dimensional failure and mixed simplices is that the parts of the boundary are only $\Nv-1$ dimensional and, therefore, the corresponding integration nodes must be used.

\footnotesize

\bibliographystyle{asa}
\bibliography{bibliography}

\normalsize

\end{document}